\documentclass[twocol]{ametsocv6.1}

\usepackage{epstopdf}
\usepackage{hyperref}
\hypersetup{colorlinks,allcolors=black}

\title{Gravity-induced collisions of uncharged cloud droplets in an electric field}

\authors{Pijush Patra \thanks{Pijush Patra's current affiliation: 
Nordita, KTH Royal Institute of Technology and Stockholm University, Stockholm 10691, Sweden.}and Anubhab Roy \correspondingauthor{Anubhab Roy, anubhab@iitm.ac.in}}
\affiliation{Department of Applied Mechanics, Indian Institute of Technology Madras, Chennai, Tamil nadu 600036, India}

\abstract{We investigate the collisions of uncharged, conducting droplets settling under gravity in the presence of an external electric field. Previous studies have derived a near-field asymptotic expression for the electric-field-induced attraction, suggesting that this force can overcome lubrication resistance and drive surface-to-surface contact between two spherical conductors within a finite time. However, for droplets moving in air, traditional lubrication theory breaks down when the inter-droplet gap approaches the mean free path of air molecules. To account for this, we incorporate non-continuum hydrodynamic effects to estimate the gravity-driven collision efficiency under electric-field-induced forces. This study examines how an external electric field influences the trajectories of settling droplet pairs of unequal sizes. By analyzing their motion, we compute collision efficiencies and explore their dependence on droplet size ratio, electric field strength, the angle between the field and gravity, and key dimensionless parameters governing electric-field-induced and van der Waals forces. Our findings reveal that electric-field-induced forces significantly enhance collision efficiency, highlighting their critical role in droplet coalescence dynamics.}

\begin{document}
\maketitle




\section{Introduction}\label{Introduction}
The initiation of warm rain has long been a subject of interest in the cloud microphysics community, with a central question being: what mechanisms govern the evolution of droplet size distributions (DSD) in atmospheric clouds \citep{pruppacher1997microphysics,devenish2012droplet}? Growth through condensation alone is unlikely to produce the broad droplet size spectrum observed in in situ measurements \citep{prabha2011microphysics,khain2013mechanism}. Consequently, collisions and subsequent coalescence between droplets are the primary drivers of rain formation in warm clouds, with DSD evolution heavily dependent on the collision rate. Extensive research has examined the role of turbulence and gravitational settling in droplet collisions (see \citealt{shaw2003particle,grabowski2013growth} and references therein). However, relatively few studies have investigated the influence of electrostatic forces on these interactions \citep{schlamp1976numerical,tinsley2000effects,khain2004rain,patra2023collision,dubey2024critical}. Incorporating an accurate parameterization of electrostatic forces into large-scale models, such as numerical weather prediction simulations, could enhance forecasting accuracy. More importantly, electrostatic forces may play a crucial role in initiating collisions among small droplets, introducing size disparities necessary for further growth. This study focuses on droplet collisions driven by electrostatic forces arising from an external electric field.

Most clouds are naturally electrified, making electrostatic interactions between cloud droplets a significant factor in the collision-coalescence process \citep{pruppacher1997microphysics,wang2013physics}. A vertically downward fair-weather electric field exists due to the potential difference between the Earth's surface and the upper atmosphere. In thunderclouds, this electric field intensifies rapidly due to charge separation, driven by various charging mechanisms broadly categorized as convective, inductive, and non-inductive \citep{kamra1975role,latham1981electrification,williams1989relationship,saunders1993review,mareev2017role}. Detailed discussions on these mechanisms can be found in Sec. 18.5 of \citet{pruppacher1997microphysics} and Sec. 14.4 of \citet{wang2013physics}. Laboratory experiments have consistently demonstrated that collisional charging during ice-ice collisions plays a crucial role in thundercloud electrification \citep{reynolds1957thunderstorm,latham1965electrification,takahashi1978riming,gaskell1980charge,jayaratne1983laboratory,mason2000charge,saunders2006laboratory,turner2022effects}. These findings have inspired the development of several theoretical models for collisional charging \citep{baker1989charge,baker1994mechanism,dash2001theory,dash2003surface,jungwirth2005possible,kang2023new}.  

The strength of the electric field in fair-weather clouds typically ranges from \(10^2\) to \(10^3\) V/m. However, field measurements (see \citet{gunn1948electric}; Chap. 18, pp. 804-811 of \citet{pruppacher1997microphysics}; Chap. 3, pp. 86-87 of \citet{rakov2003lightning}; \citet{trinh2020determining}) indicate that electric fields in highly electrified clouds can reach magnitudes of \(O(10^4 - 10^5)\) V/m. \citet{winn1974measurements} even reported values as high as \(4 \times 10^5\) V/m. Notably, these field studies suggest that the electric field is not always directed vertically downward; it can also act horizontally or at an oblique angle relative to gravity.  

Strong electric fields in thunderclouds can significantly influence droplet collision dynamics. Recent field observations by \citet{mudiar2018quantification,mudiar2021electric} suggest that intense electric fields in highly electrified clouds enhance raindrop growth and increase rainfall rates. Additionally, a series of wind tunnel experiments by Kamra and co-workers demonstrate that strong electric fields substantially alter the microphysical properties of electrified clouds, particularly affecting binary collisions between water droplets by modifying their impact velocities and deformation characteristics \citep{bhalwankar2007wind,bhalwankar2009wind,bhalwankar2023binary,pawar2024effects}.

An applied electric field, no matter how weak, induces electric charges of opposite signs on the nearest sides of two uncharged conducting droplets. As the droplets move closer, the mutual interactions between these induced charges enhance the local electric field in the region between the two droplets. This locally amplified electric field surpasses the strength of the imposed field, resulting in an increase in electrostatic attraction force as the droplets approach each other \citep{davis1964two}. Notably, this electric-field-induced attractive force between two uncharged conducting droplets increases without bound as their separation approaches zero \citep{lekner2011near,lekner2013forces}. As a result, this force could overcome lubrication resistance, facilitating surface-to-surface contact between droplets. However, it is important to note that the electric-field-induced force diminishes significantly at larger droplet separation distances. Thus, two widely separated cloud droplets subject to an external electric field are unlikely to make contact unless influenced by a background flow, gravity, or thermal fluctuations that drive them into closer proximity.

The calculation of electrostatic force between two uncharged or charged spherical droplets subject to an external electric field has been thoroughly investigated. One of the pioneering studies on this topic was carried out by \citet{davis1964two}, who calculated the electrostatic forces on two charged spherical conductors subject to an external electric field by integrating the electrical stresses over their surfaces. This electric stress is directly related to the gradient of the electric potential $\varPhi_{\text{el}}$. \citet{davis1964two} determined the variation of the potential field by solving the Laplace equation for $\varPhi_{\text{el}}$ in a bispherical coordinate system. Finally, \citet{davis1964two} derived the expressions for forces in terms of sphere sizes, surface charges on the spheres, the magnitude of the imposed electric field, the angle between the line joining the centers and the direction of the electric field, and the force coefficients $F_i$ ($i=1,2,\cdot\cdot\cdot,10$) that depend on the relative geometry of the two spheres. The coefficients $F_5$, $F_6$, and $F_7$ are required to describe the forces without an imposed electric field. Whereas $F_1$, $F_2$, and $F_8$ are required to calculate the forces between uncharged spherical conductors in an external electric field. The coefficients $F_3$, $F_4$, $F_9$, and $F_{10}$ capture the coupled effects of surface charges and the imposed electric field. The expressions of these force coefficients involve infinite series summations. The convergence of these series becomes extremely slow for small separation distances. To avoid this issue, \citet{lekner2013forces} calculated the electric-field-induced forces using the energy method, where he wrote the electrostatic energy of the system in terms of a polarizability tensor. Longitudinal and transverse polarizabilities are sufficient to describe the system energy and forces in the case of a two-sphere system. To obtain the electric-field-induced forces in the close approach of two arbitrary-sized spherical conductors, we utilize the work of \citet{lekner2011polarizability}, who derived the exact analytical expressions for the longitudinal and transverse polarizabilities for small separation distances. The forces acting along the line joining two centers are equal and opposite and depend on the partial derivatives of these polarizabilities with respect to the separation distance. The forces normal to the line of centers produce torque on the two-sphere system, which depends on the difference between the longitudinal and transverse polarizabilities. This torque always acts to align the line of centers with the direction of the external electric field.

Droplet volume fractions in atmospheric clouds are low [about $O(10^{-6})$] \citep{grabowski2013growth}, and therefore, we consider only binary collisions. The rate equation for the droplet number density when two species are present is:
\begin{gather}
-\frac{d n_1}{d t} = -\frac{d n_2}{d t} = K_{12},
\label{Number_density_equation}
\end{gather} 
where $K_{12}$ is the collision rate between droplet categories of number densities $n_1$ and $n_2$. \citet{smoluchowski1918versuch} studied the ideal collision rate $K_{12}^0$ for two non-interacting spheres settling under gravity in a quiescent fluid and found that $K_{12}^0=n_1 n_2 [2\pi (\rho_p-\rho_f)(a_1^2-a_2^2)g(a_1+a_2)^2]/(9\mu_f)$, where $a_1$ is the radius of the larger droplet and, $a_2$ is that of the smaller droplet, $\rho_p$ and $\rho_f$ are the droplet and fluid densities, $g$ is the acceleration due to gravity, and $\mu_f$ is the dynamic viscosity of the fluid. The theoretical study of droplet collision rate with hydrodynamic and nonhydrodynamic interactions is a challenging task. These interactions alter the collision rate by modifying the relative velocity between the droplet pair at close separations. The collision efficiency $E_{12}=K_{12}/K_{12}^0$, which is the ratio of the collision rate with interactions to that obtained ignoring interactions (i.e., the ideal collision rate), captures the effects of interactions on the collision rate. \citet{davis1984rate} and \citet{melik1984gravity} predicted the collision efficiency for two unequal-sized rigid spheres sedimenting due to gravity and interacting via continuum hydrodynamics and van der Waals (vdW) forces. \citet{zhang1991rate} and \citet{rother2022gravitational,ababaei2023collision} calculated the collision efficiency induced by interfacial mobilities for two differentially sedimenting viscous drops without and with inertial effects. Previous studies have utilized the work of \citet{davis1964two} to analyze the collision rate of uncharged or charged sedimenting cloud droplets in an external electric field \citep{sartor1960some,plumlee1965cloud,semonin1966collision,schlamp1976numerical,guo2021enhancement}. These studies do not consider the exact hydrodynamics for small separation distances between the droplets (i.e., lubrication interactions). Considering the internal circulation of fluid droplets, \citet{zhang1995theoretical} predicted the electric field-enhanced collision rate of two uncharged conducting spherical droplets settling under gravity. However, these works did not account for the exact analytical form of electric-field-induced forces in the lubrication regime. Recently, \citet{thiruvenkadam2023pair} analyzed relative trajectories of two arbitrary-sized uncharged conducting spheres without gravity and showed that because of the divergent nature of the electric-field-induced forces in the lubrication region, spheres could come into contact in a finite time. Motivated by this, we examine the effects of electric-field-induced forces on pair trajectories and the collision rate of two uncharged conducting droplets sedimenting in still air.

The characteristic hydrodynamic and electric stresses for a spherical water droplet sedimenting in air and subject to an electric field are $\mu_f U_s/a$ and $\epsilon_0 E_0^2$, respectively, where $a$ is the radius of the droplet, $U_s$ is the Hadamard–Rybczynski settling speed, $\epsilon_0$ is the permittivity of air, and $E_0$ is the magnitude of the external electric field. Here, $U_s = 2\left((\rho_p-\rho_f) g a^2/3\mu_f\right)(\hat{\mu}+1)/(3\hat{\mu}+2)$, where $\hat{\mu}$ is the droplet-to-medium viscosity ratio. The capillary number, $Ca = \mu_f U_s/\gamma$, where $\gamma$ is the surface tension at the air-water interface, measures the relative strength of hydrodynamic and capillary stresses. The electric stress to surface tension stress ratio defines the electric capillary number $Ca_E = \epsilon E_0^2 a/\gamma$. The droplet shape depends on the dimensionless quantities $Ca$ and $Ca_E$, and one can assume that the droplet will remain spherical if $Ca, Ca_E \ll 1$. Let us consider a water droplet in air with $a = 10$ \textmu m, $\rho_p \approx 10^3$ kg m$^{-3}$, $\rho_f \approx 1$ kg m$^{-3}$, $\mu_f \approx 1.8 \times 10^{-5}$ Pa.s, $\hat{\mu} \approx 10^2$, $E_0 = 10^5$ Vm$^{-1}$, $\epsilon = \epsilon_0 = 8.85 \times 10^{-12}$ Fm$^{-1}$, and $\gamma \approx 72 \times 10^{-3}$ Nm$^{-1}$. Using these data, we find that $Ca \approx 3 \times 10^{-6}$ and $Ca_E \approx 1.2 \times 10^{-5}$ are sufficiently small, and thus, we can safely neglect droplet deformation. Furthermore, as $\hat{\mu} = O(10^2)$ for water droplets in air, the mobility of droplet interfaces is insignificant, and the small droplets will almost behave like rigid spherical particles.

We ignore the role of flow inertia and droplet inertia in collision dynamics. We define the Reynolds number based on the terminal settling speed and radius of the larger droplet, i.e., $Re_p=[2\rho_f(\rho_p-\rho_f) g a_1^3]/(9\mu_f^2)$. The Stokes number defined as $St=[16\rho_p(\rho_p-\rho_f) g (a_1^2-a_2^2)(a_1 a_2)^{3/2}]/[81\mu_f^2(a_1+a_2)^2]$ \citep{davis1984rate} captures the droplet inertia. The Peclet number $Pe=2\pi (\rho_p-\rho_f)a_1^4 \kappa (1-\kappa^2)g/(3 k_B T)$ measures the relative importance of gravitational advection and diffusion due to thermal fluctuations \citep{zinchenko1994gravity}, where $\kappa=a_2/a_1 < 1$ is the size ratio of the droplet pair. Here, $k_B=1.318 \times 10^{-23}$JK$^{-1}$ is the Boltzmann's constant, and $T$ is the absolute temperature. Let us calculate typical values of $Re_p$, $St$, and $Pe$ for water droplets in a warm cloud at $T=275$ K when $a_1 = 10$ \textmu m. We found $Re_p \approx 0.007$ (negligible fluid inertia); $St \approx 0.54$ for $\kappa = 0.3$ and $St \approx 0.03$ for $\kappa = 0.99$; $Pe \approx 1504$ for $\kappa = 0.3$ and $Pe \approx 1085$ for $\kappa=0.99$. The above representative values of $St$ suggest that the particle inertia is negligible for a nearly equal-sized droplet pair of radii less than $15$ \textmu m. In warm cumulus clouds, condensation is the dominant growth mechanism for droplets of radii up to $15$ \textmu m. Thus, condensation leads to a nearly monodisperse size distribution. So, the negligible particle inertia assumption is valid at the lower end of the size gap of $15-40$ \textmu m droplets. Our non-inertial collision calculations will work as reference calculations for future studies in this area. The relative velocity between two non-interacting spherical droplets settling under gravity in a quiescent environment is given by $\boldsymbol{V}^{(0)}_{12} = 2 (\rho_p-\rho_f) \left(a^2_1-a^2_2\right) \textit{\textbf{g}}/(9 \mu_f)$. The collisional time scale of two differentially sedimenting droplets would be $\tau_{\text{coll}} = a^*/|\boldsymbol{V}^{(0)}_{12}|$, where $a^*=(a_1+a_2)/2$ is the average radius of the two droplets. For water droplets in air, $\tau_{\text{coll}} \approx 8 \times 10^{-4}$ s when $a_1 = 10$ \textmu m and $a_2 = 5$ \textmu m. The droplet polarization time scale $\tau = \epsilon_0/\sigma \approx 6.5$ min, where $\sigma \approx 2.3 \times 10^{-14}$ Sm$^{-1}$ is the conductivity of clear air at sea level. Air conductivity inside a warm cloud can range from $1/40$ to $1/3$ of the fair-weather sea level conductivity \citep{pruppacher1997microphysics}, which suggests that $\tau$ can vary approximately from $20$ min to $2$ h. Since $\tau_{\text{coll}} \ll \tau$, we can neglect the effects of surface charge convection on the collision dynamics.

When the separation distance between two droplets is small to moderate, they disturb the velocity field around each other. These disturbances give rise to additional hydrodynamic resistance on each droplet. This way, droplets interact with each other through the fluid medium. These hydrodynamic interactions between the droplets can significantly modulate the collision dynamics. Hydrodynamics interactions between a pair of droplets in Stokes flow conditions are well studied (see \citealt{guazzelli2011physical,kim2013microhydrodynamics}). This hydrodynamic resistance is $O(1/f(\xi))$ in the lubrication region, where $\xi = (r-(a_1+a_2))/a^* = (r/a^*)-2$ is the dimensionless (nondimensionalized by $a^*$) surface-to-surface distance distance between the two droplets. The function $f(\xi) = \xi$ for two rigid spherical particles with continuum hydrodynamic interactions \citep{batchelor1972hydrodynamic} and $f(\xi) = \sqrt{\xi}$ for two spherical viscous drops interacting via continuum hydrodynamics \citep{davis1989lubrication}. For particles interacting in a gaseous medium, the continuum lubrication approximation would no longer be valid when the gap thickness between two surfaces is less than the mean free path of the medium $\lambda_0$, and then one needs to consider non-continuum lubrication resistance where $f(\xi) =\ln(\ln(Kn/\xi))/Kn$ \citep{sundar96non}. Here, $Kn$ is the Knudsen number that captures the significance of non-continuum interactions and is defined as the ratio of the mean free path of the medium to the mean radius of the droplets. Previous studies have obtained collision rates due to non-continuum interactions for droplets subject to Brownian motion \citep{patra2022brownian}, differential sedimentation and uniaxial compressional flow \citep{dhanasekaran2021collision}, simple shear flow \citep{patra_koch_roy_2022}, and turbulent flow \citep{dhanasekaran2021turbulent}.

We organize the paper as follows. In Sec. \ref{Problem_formulation}, we will formulate the problem and outline the procedure for calculating the collision rate and efficiency using trajectory analysis. Then, in Sec. \ref{Results_and_discussion}, we will calculate the collision efficiency of a pair of hydrodynamically interacting droplets due to the combined effect of gravity, electric field, and van der Waals forces. Finally, in Sec. \ref{Conclusions}, we will summarize our results and discuss their implications.

\section{Problem formulation}\label{Problem_formulation}
\subsection{The relative velocity between a pair of droplets}\label{The_relative_velocity_between_a_pair_of_droplets}
We consider two uncharged, conducting cloud droplets settling under gravity while subjected to a uniform external electric field inclined at an angle \(\eta\) relative to the gravitational direction. Hydrodynamic interactions between the droplets are also accounted for. In dilute systems like clouds, the probability of a third droplet significantly influencing the relative motion of two interacting droplets is negligible. Therefore, our analysis focuses on binary collisions.  

\begin{figure}
\centering
\includegraphics[width=0.4\textwidth]{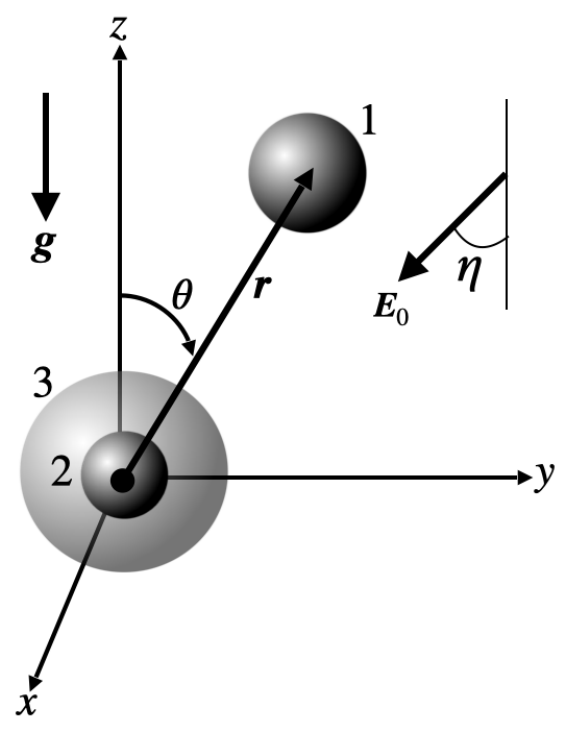}
\caption{Schematic of the coordinate system used in the analysis. `1' indicates the sphere with radius $a_1$ and; `2' indicates the sphere with radius $a_2$. The sphere marked `3' is the collision sphere of radius $a_1+a_2$. The electric field $\textit{\textbf{E}}_0$ makes an angle $\eta$ with the gravity direction. We denote $\hat{e}_r$, $\hat{e}_{\theta}$ and $\hat{e}_{\phi}$ the unit vectors in $r$, $\theta$ and $\phi$ directions respectively.}
\label{Schematic_binary_collisions_gravity_electric_field}
\end{figure}

To describe the relative motion, we track the velocity of a satellite droplet (radius \(a_1\)) with respect to a test droplet (radius \(a_2\)) (see Figure \ref{Schematic_binary_collisions_gravity_electric_field}). The relative velocity between the droplets is given by \(\boldsymbol{V}_{12} = \boldsymbol{V}_{1} - \boldsymbol{V}_{2}\). Since the fluid motion generated by a settling droplet pair is assumed to be slow, the surrounding disturbance flow field satisfies the Stokes equations for creeping flow. The linearity of these equations allows us to express the resultant relative velocity as a vector sum of contributions from gravity, electric-field-induced forces, and van der Waals forces \citep{batchelor1976brownian,batchelor1982sedimentation,davis1984rate,zhang1995theoretical}:  

\begin{multline}
\boldsymbol{V}_{12}(\boldsymbol{r}) = \boldsymbol{V}^{(0)}_{12} \mathbf{\cdot} \left[ L\frac{\boldsymbol{rr}}{r^2} + M\left(\textbf{\textit{I}} - \frac{\boldsymbol{rr}}{r^2} \right)\right] + \frac{1}{6\pi \mu_f } \left(\frac{1}{a_1} + \frac{1}{a_2}\right) \\
\left[ G\frac{\boldsymbol{rr}}{r^2} + H\left(\textbf{\textit{I}} - \frac{\boldsymbol{rr}}{r^2}\right)\right] \mathbf{\cdot} \left(\boldsymbol{F}_E + \boldsymbol{F}_{\text{vdW}}\right),
\label{general_relative_velocity_equation}
\end{multline}  

where \(\boldsymbol{r}\) is the vector from the center of droplet 2 (test droplet) to the center of droplet 1 (satellite droplet), and \(r = |\boldsymbol{r}|\). Here, \(\textbf{\textit{I}}\) denotes the unit second-order tensor, while \(\boldsymbol{F}_E\) and \(\boldsymbol{F}_{\text{vdW}}\) represent the electric-field-induced and van der Waals forces, respectively.  

The mobility functions \(L\) and \(M\) characterize the relative motion of two settling droplets under gravity, with \(L\) governing motion along the line of centers and \(M\) governing motion perpendicular to it \citep{batchelor1982sedimentation}. Similarly, \(G\) and \(H\) represent the corresponding mobility functions for droplets influenced by nonhydrodynamic forces. These functions depend on the droplet size ratio \(\kappa = a_2/a_1\) and the dimensionless center-to-center distance \(r/a^*\).  To compute these mobility functions, we leverage the work of \citet{wang1994collision} and \citet{zinchenko1994gravity}, who determined axisymmetric mobilities for continuum hydrodynamic interactions by solving the Stokes equations in a bispherical coordinate system. Additionally, we use the twin multipole expansion method developed by \citet{jeffrey1984calculation} to determine asymmetric mobilities. However, as the separation between droplet surfaces decreases, these series solutions require an increasingly large number of terms to achieve convergence, making them computationally inefficient. When the gap between the droplets becomes small, lubrication approximations for the mobility functions, available in existing literature \citep{batchelor1982sedimentation,jeffrey1984calculation}, provide a more efficient alternative.  

In a continuum framework, hydrodynamic interactions prevent surface-to-surface contact between two rigid spheres within a finite time unless an attractive nonhydrodynamic force—such as the van der Waals force—is present. However, even in the absence of such forces, non-continuum lubrication effects can enable collisions to occur in finite time. This arises because the resistance functions associated with motion along the line connecting the two droplet centers exhibit a weaker divergence rate in the non-continuum regime \citep{dhanasekaran2021collision,patra2022brownian}.\citet{sundar96non} derived expressions for non-continuum lubrication forces for axisymmetric motion. Building on this work, \citet{dhanasekaran2021collision} recently incorporated non-continuum lubrication effects into the axisymmetric mobility functions. Their approach treats lubrication interactions as continuum for \(\xi > O(Kn)\) and non-continuum for \(\xi \leq O(Kn)\), where \(\xi\) is the dimensionless gap between the two spheres. In this study, we adopt the uniformly valid solutions for \(L\) and \(G\) developed by \citet{dhanasekaran2021collision}. For asymmetric mobilities, \(M\) and \(H\), continuum hydrodynamics provides a reasonable approximation at all separations because these mobility functions approach a finite value as \(\xi \to 0\). Consequently, continuum breakdown is not expected to significantly affect asymmetric relative motions for an inertialess droplet pair, allowing us to use continuum hydrodynamics for asymmetric mobilities throughout. However, when calculating the collision rate for an inertial droplet pair in a gaseous medium, non-continuum lubrication effects must also be considered for relative motions in the tangential directions \citep{how2021non}. 

The calculation of electrostatic forces between two conducting droplets in an external electric field is primarily a boundary value problem where one needs to solve Laplace's equation for the potential field in a bispherical coordinate system \citep{davis1964two}. Once the potential field is determined, the electric-field-induced forces on each droplet can be obtained by integrating the electrical stresses over their surfaces. In this case, the forces acting on the two droplets are equal and opposite. Given the axisymmetric nature of the problem, the force on each droplet can be decomposed into two components: one directed along the line of centers and the other perpendicular to it. The expressions for the electric-field-induced forces in $r$ and $\theta$ directions are given by \cite{davis1964two}:
\begin{gather}
F_E^r = -4\pi\epsilon_0a_2^2E_0^2\left(F_1\cos^2(\theta-\eta)+F_2\sin^2(\theta-\eta)\right), \label{F_E^r} \\
F_E^{\theta} = 4\pi\epsilon_0a_2^2E_0^2 F_8\sin 2(\theta-\eta), \label{F_E^theta}
\end{gather}
where $\theta-\eta$ is the angle between the electric field vector $\boldsymbol{E}_0$ and the line joining the centers of two spheres, and as discussed in \S \ref{Introduction}, $F_1, F_2, F_8$ are the force coefficients that depend on the center-to-center distance and the size ratio of the two droplets. The analytical expressions for these force coefficients in the near and far fields are provided in \citet{thiruvenkadam2023pair}.

The van der Waals attraction force always acts along the line joining the centers of the two droplets, and therefore we can write $F_{\text{vdW}}=-d \varPhi_{\text{vdW}}/d r$, where $\varPhi_{\text{vdW}}$ is the van der Waals potential. Most previous collision calculations used the unretarded form of $\varPhi_{\text{vdW}}$ derived by Hamaker \cite{hamaker1937london}, who obtained an analytical expression for $\varPhi_{\text{vdW}}$ using a pairwise additivity theory. However, Hamaker's analysis did not consider the effects of retardation due to the finite propagation speed of electromagnetic waves. One must consider the effects of retardation when separation is comparable to or more than the London wavelength $\lambda_L (\approx 0.1$ \textmu m). In the present analysis, we use the work of \citet{zinchenko1994gravity}, who obtained the retarded van der Waals potential by integrating the dispersion energy between two molecules. The functional form of $\varPhi_{\text{vdW}}$ depends on the dimensionless center-to-center distance, the size ratio, the nondimensional quantity $N_L$ defined as the radius of the droplets scaled with $\lambda_L$ (i.e., $N_L = 2 \pi \left(a_1+a_2\right)/\lambda_L = 2 \pi a_1 \left(1+\kappa\right)/\lambda_L$), and the Hamaker constant $A_H$.

\begin{figure}
\centering
\includegraphics[width=0.48\textwidth]{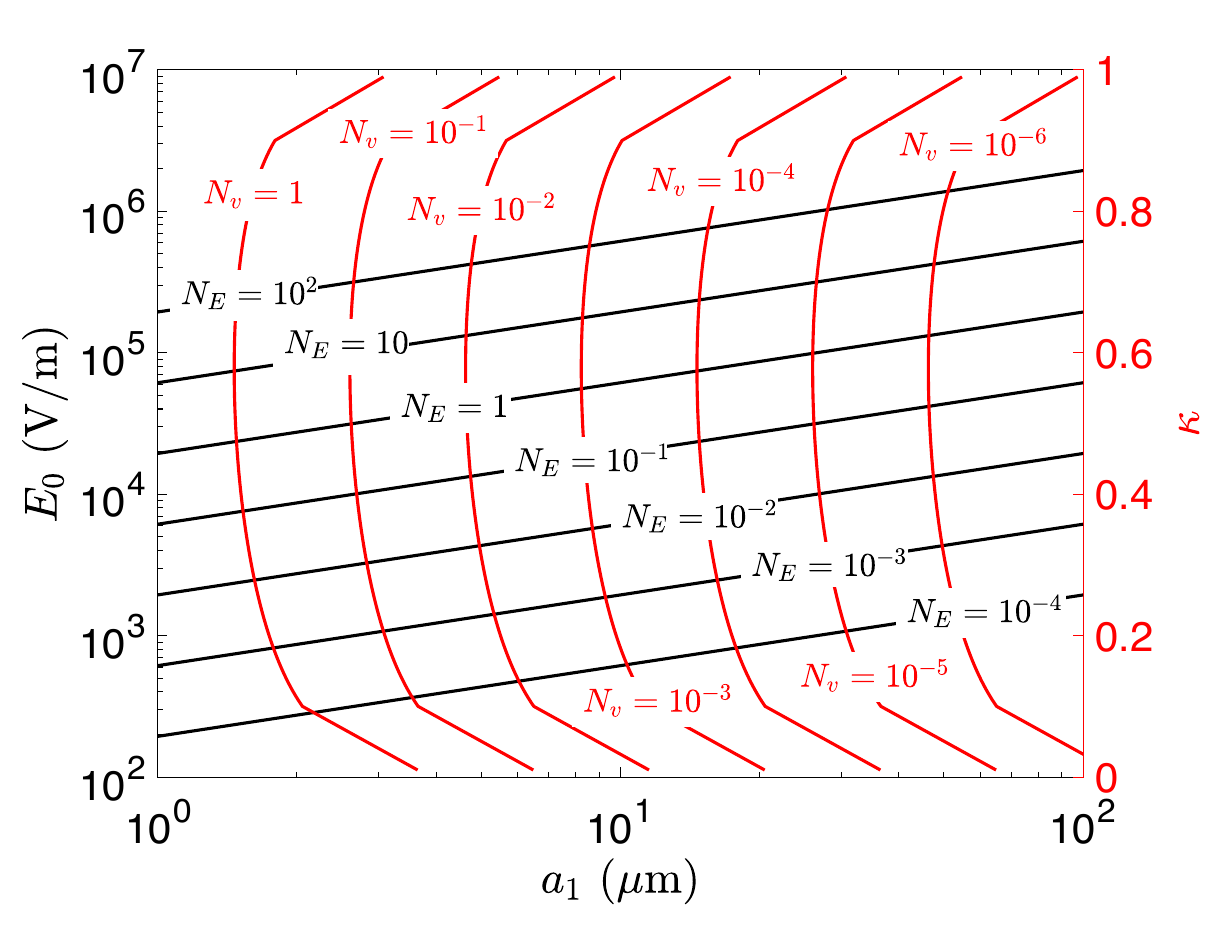}
\caption{Typical sizes of droplets and electric fields in clouds. The black and red lines correspond to constant $N_E$ and $N_v$, respectively}
\label{NE_Nv_parameter values_in_a1_kappa_E0_space}
\end{figure}

We choose a spherical coordinate system $(r,\theta,\phi)$ whose origin coincides with the center of the test droplet. We nondimensionalize the governing equation by considering $a^*$, $V^{(0)}_{12}=|\boldsymbol{V}^{(0)}_{12}|$ and $a^*/V^{(0)}_{12}$ as the characteristic length, velocity and time scale of the problem. From here onward, we denote the nondimensional radial separation between the centers of the two droplets by $r$. Thus, $r$ lies in the range $2$ (referred to as the collision sphere, indicated by sphere 3 in figure \ref{Schematic_binary_collisions_gravity_electric_field}) to $\infty$ (where one droplet does not influence the other). Similarly, we scale the coordinates by $a^*$ and denote them with an overbar (i.e., $\overline{x} = x/a^*$, $\overline{y} = y/a^*$, and $\overline{z} = z/a^*$ are dimensionless coordinates). The size ratio $\kappa$ that captures the geometry of the two-droplet system can vary in the range $(0,1]$. The dimensionless relative velocity $\boldsymbol{v} = \boldsymbol{V}_{12}/V^{(0)}_{12}$ can be written as $\boldsymbol{v} = v_r \hat{e}_r + v_{\theta} \hat{e}_{\theta} + v_{\phi} \hat{e}_{\phi}$, where
    
\begin{gather}
v_r = \frac{d r}{d t} = -L\cos\theta - N_E G \left(\frac{\kappa}{1-\kappa}\right)\Big(F_1\cos^2(\theta-\eta) \nonumber \\  +F_2\sin^2(\theta-\eta)\Big) - N_v G \frac{d \varPhi_{\text{vdW}}}{d r},
\label{vr_equation} \\
v_{\theta} = r\frac{d \theta}{d t} = M \sin\theta + N_E H \left(\frac{\kappa}{1-\kappa}\right) F_8\sin 2(\theta-\eta),
\label{vtheta_equation} \\
v_{\phi} = 0, \label{vphi_equation}
\end{gather}

with $N_E$ and $N_v$ are dimensionless quantities that capture the relative strength of electric-field-induced and retarded van der Waals forces to gravity:
\begin{gather}
N_E = \frac{3\epsilon_0 E_0^2}{(\rho_p-\rho_f)a_1g},
\label{expression_for_NE} \\
N_v = \frac{3 A_H}{2\pi\kappa\left(1-\kappa^2\right)(\rho_p-\rho_f) g a_1^4}. \label{expression_for_NF}    
\end{gather}
In this problem, $N_E$ primarily depends on the strength of the electric field. We intentionally excluded the size ratio term from the definition of $N_E$ to ensure that the dependence of the collision dynamics on $N_E$ directly correlates with its dependence on $E_0$. The parameter $N_v$ depends on the size ratio, and this definition is consistent with the earlier works that studied the effect of van der Waals force on the collisions of droplets settling under gravity. In figure \ref{NE_Nv_parameter values_in_a1_kappa_E0_space}, we present typical sizes of cloud droplets and electric fields in clouds, along with estimates of the parameters $N_E$ and $N_v$ shown using black and red contours.

\subsection{Expressions for the collision rate and efficiency}\label{Expressions_for_the_collision_rate_and_efficiency}
Mathematically, the collision rate $K_{12}$ is equal to the flux of droplets into the collision sphere of nondimensional radius $r=2$ and can be expressed in terms of the pair distribution function $P(r)$ and the droplet relative velocity $\boldsymbol{v}$ by
\begin{gather}
K_{12} = -n_1n_2 V^{(0)}_{12}\left(a^*\right)^2 \int_{(r=2)\&\left(\boldsymbol{v}\mathbf{\cdot}\textbf{\textit{n}}<0\right)}  \left(\boldsymbol{v}\mathbf{\cdot}\textbf{\textit{n}}\right) P dA,
\label{Collision_rate_general_dimensional}
\end{gather}
where $\textbf{\textit{n}}$ is the outward unit normal at the collision sphere. The condition $\boldsymbol{v}\mathbf{\cdot}\textbf{\textit{n}}<0$ in (\ref{Collision_rate_general_dimensional}) implies that the radial relative velocity must be inward at all separations for two droplets coming into contact. For a dilute system like clouds, the pair-distribution function is governed by the quasi-steady Fokker-Planck equation for the region of space outside the collision sphere:
\begin{gather}
\boldsymbol{\nabla}\mathbf{\cdot}\left(P\boldsymbol{v}\right) = 0.
\label{Probability_conservation}
\end{gather}
The uncorrelated motion of the droplets in the far field defines the boundary condition: $P \rightarrow 1$ as $r \rightarrow \infty$. For the calculation purpose, we take $r=r_{\infty}$, which is large but finite.

For negligible thermal fluctuations (i.e., Pe $\gg 1$), the relative motion between two droplets induced by gravity and an external electric field is deterministic, and therefore, we can find the collision rate using the trajectory analysis. Using equation (\ref{Probability_conservation}) and the divergence theorem, the integral in (\ref{Collision_rate_general_dimensional}) can be taken over the surface that encloses the volume occupied by all trajectories that originate at $r=r_{\infty}$ and terminate at $r=2$. Thus, the flux through the cross-section $A_c$ of this volume at $r=r_{\infty}$ defines the collision rate. We call $A_c$ the upstream interception area. At large separations, the electric-field-induced and van der Waals forces become insignificant, and gravity solely drives the relative motion between two unequal-sized droplets. Now, at $r=r_{\infty}$, the pair distribution function $P=1$ and $\boldsymbol{v}\mathbf{\cdot}\textbf{\textit{n}}'=-1$, where $\textbf{\textit{n}}'$ is the unit outward normal vector at the area element of $A_c$. Therefore, the expression for the collision rate becomes
\begin{gather}
K_{12} = -n_1n_2 V^{(0)}_{12}\left(a^*\right)^2\int_{A_c}\left(\boldsymbol{v}\mathbf{\cdot}\textbf{\textit{n}}'\right)|_{r_{\infty}} P dA \nonumber \\ = \frac{1}{4}n_1n_2V^{(0)}_{12}\left(a_1+a_2\right)^2\pi \left(\dfrac{\overline{y}_{c+}^2}{2}+\dfrac{\overline{y}_{c-}^2}{2}\right),
\label{Collision_rate_upstream_interception_area}    
\end{gather}    
where $\overline{y}_{c\pm}=y_{c\pm}/a^*$ ($y_{c+}$ and $y_{c-}$ are the dimensional critical impact parameters for the positive and negative side of $y-$axis, respectively) are the dimensionless critical impact parameters which define the nondimensional radii of the upstream collisional semi-circles at $r=r_{\infty}$. In other words, these critical impact parameters are the largest possible horizontal distances from the gravity axis at the far-field for which two widely separated droplets eventually collide. The relative droplet trajectory, in this case, is called the limiting colliding trajectory. Equation (\ref{Collision_rate_upstream_interception_area}) bypasses the calculation for the evolution of $P$. The collision rate $K_{12}^{0}$ without any interactions is given by the classical Smoluchowski model \citep{smoluchowski1918versuch}, where $y_{c+}=y_{c-}=a_1+a_2$ (i.e., $\overline{y}_{c+}=\overline{y}_{c-}=2$). Thus, the expression for the ideal collision rate becomes
\begin{gather}
K_{12}^{0} = n_1n_2V^{(0)}_{12}\pi \left(a_1+a_2\right)^2,
\label{ideal_collision_rate}
\end{gather}
The collision efficiency $E_{12}$ is defined as the ratio of $K_{12}$ to $K_{12}^{0}$:
\begin{gather}
E_{12} = \frac{K_{12}}{K_{12}^{0}} = \frac{1}{4}\left(\dfrac{\overline{y}_{c+}^2}{2}+\dfrac{\overline{y}_{c-}^2}{2}\right).
\label{Collision_efficiency}
\end{gather}
The droplet pair collides when the initial $\overline{y}$ values at the far-field belong to $[\overline{y}_{c-}$,$\overline{y}_{c+}]$. Therefore, the problem now becomes one of determining the dimensionless critical impact parameters $\overline{y}_{c\pm}$. We find the limiting colliding trajectories by integrating the following dimensionless trajectory equation
\begin{gather}
\frac{d\theta}{dr}=\frac{v_{\theta}}{r v_r},
\label{final_trajectory_equation}
\end{gather}
The above equation describes the relative trajectory of a particle pair due to the combined effects of gravity, hydrodynamic interactions, electric-field-induced forces, and van der Waals forces. We calculate the collision rate and efficiency from equations (\ref{Collision_rate_upstream_interception_area}) and (\ref{Collision_efficiency}) after determining the critical impact parameters using trajectory analysis. Starting with different appropriate initial conditions, we obtain pair trajectories by integrating (\ref{final_trajectory_equation}). Out of these trajectories, the limiting colliding trajectories are those beyond which one droplet moves past the other without touching. The closed-form analytical expression for the collision efficiency in the absence of non-hydrodynamic forces (i.e., $N_E = N_v =0$ in this case) is given by \citep{davis1984rate}
\begin{gather}
E_{12}=\exp\left(-2\int_2^{\infty}\frac{M-L}{r L}dr\right).
\label{collision_efficiency_due_to_hydrodynamics_only}
\end{gather}

\section{Results and discussion}\label{Results_and_discussion}
Colliding trajectories are the paths followed by the centers of evolving satellite droplets that start far upstream and end at the collision surface. These colliding trajectories in the far field constitute the upstream interception area. The cost of computing the colliding trajectories will be huge if we choose initial conditions on a plane located far upstream and perpendicular to gravity because most trajectories starting from that plane would never reach the collision sphere. We avoid this issue by exploiting the quasi-steady nature of the relative trajectory equation, and thus, we employ backward integrations of (\ref{final_trajectory_equation}) using a fourth-order Runge-Kutta method with initial conditions on the collision sphere ($r=2$). However, exactly at $r = 2$, $v_r = 0$ since the hydrodynamic mobilities $L=0$ and $G=0$ at $r = 2$. To avoid this issue, we set initial conditions on a sphere of radius $2+\delta$, where $\delta$ is a small separation from the collision surface. We will show converged results without too much computational load when $\delta = 10^{-6}$. As stated in (\ref{Collision_rate_general_dimensional}), the radial relative velocity must be inward at the collision sphere for a colliding trajectory. Thus, we further reduce the computation by selecting only those points on the collision sphere where $v_r < 0$. It is important to mention here that depending on the computational need, sometimes we solve for $r(\theta)$ from $\dfrac{dr}{d\theta}=\dfrac{r v_r}{v_{\theta}}$ instead of solving for $\theta(r)$ from (\ref{final_trajectory_equation}).

\begin{figure}
\centering
\includegraphics[width=0.48\textwidth]{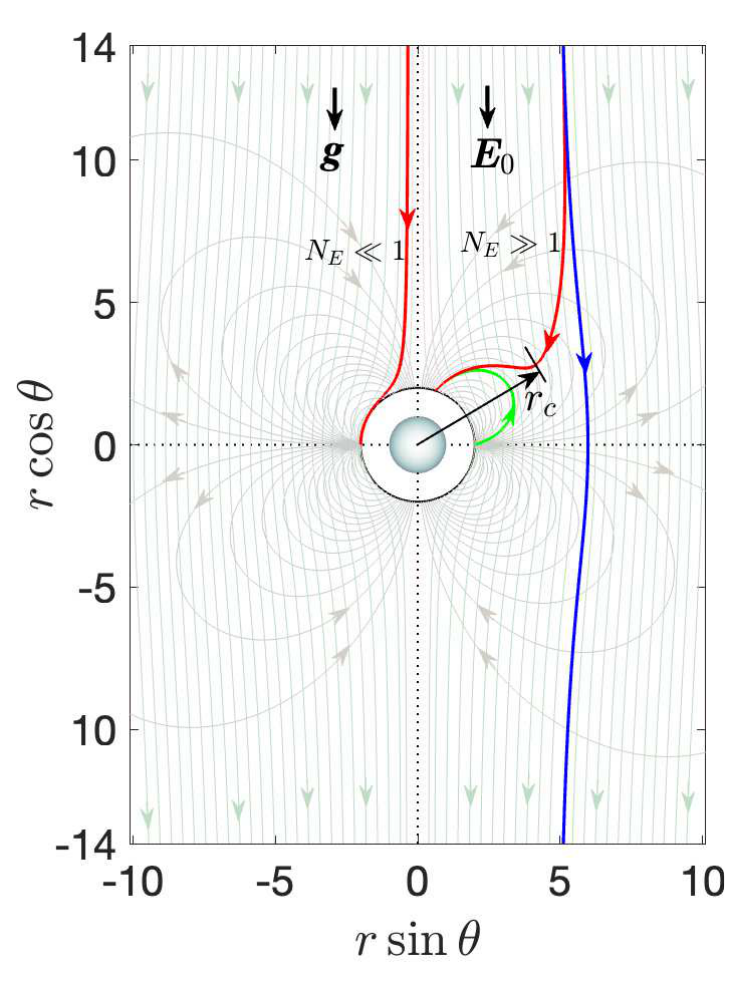}
\caption{Typical limiting colliding trajectories (continuous red lines) for weak ($N_E \ll 1$) and strong ($N_E \gg 1$) electric field. The blue and green lines indicate open and loop trajectories due to differential sedimentation and a vertical electric field only. A complete map of trajectories for gravity alone (open and colliding) and a vertical electric field alone (loop trajectories) are provided in the background for reference. $r_c$ indicates the separation distance where the effects of gravity and the electric-field-induced forces are comparable. The sphere at the center represents the test droplet, and the thin black circle represents the projection of the collision sphere. Representative arrows on the trajectories indicate their directions.}
\label{Pair_trajectories_weak_and_strong_field}
\end{figure}

\begin{figure*}
\centering
\includegraphics[width=1.0\textwidth]{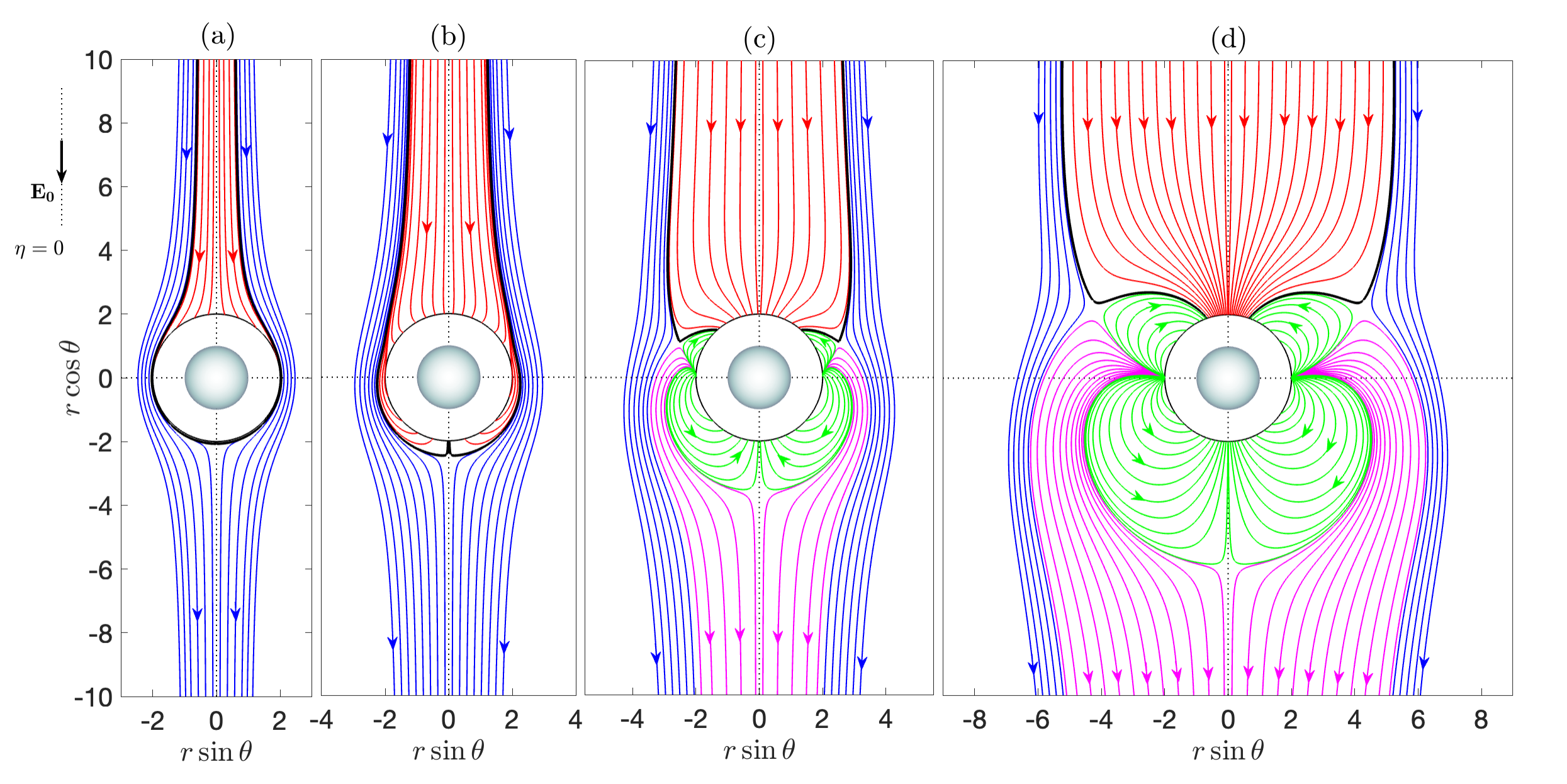}
\caption{Pair trajectories of two differentially sedimenting droplets with a vertical electric field ($\eta=0$) for non-continuum hydrodynamic interactions ($Kn=10^{-2}$) when $\kappa=0.5$, $N_v=0$, and (a) $N_E=10^{-1}$, (b) $N_E=1$, (c) $N_E=10$, and (d) $N_E=10^2$. The blue, green, red, and thick black lines are open, loop, colliding, and limiting colliding trajectories. Pink lines are a separate class of trajectory that starts from two specific locations on the collision surface and goes to infinity.}
\label{Pair_trajectories_electric_field_alpha_0}
\end{figure*}

Our primary objective in this study is to quantify how an external electric field influences the relative trajectories and collision efficiency of two conducting droplets undergoing differential sedimentation in still air. To systematically analyze this effect, we first consider a case where the droplets interact solely through hydrodynamic and electric-field-induced forces (i.e., finite \( N_E \)) while neglecting van der Waals forces (i.e., \( N_v = 0 \)). We then extend our analysis to include van der Waals interactions.  

Since the relative velocity between the droplets is independent of the azimuthal \(\phi\) coordinate, we examine the problem in a representative \( r\sin\theta - r\cos\theta \) plane. Figure \ref{Pair_trajectories_weak_and_strong_field} illustrates typical limiting colliding trajectories for weak (\( N_E \ll 1 \)) and strong (\( N_E \gg 1 \)) electric fields under conditions where \( Kn=10^{-2} \), \( \kappa=0.9 \), and \( \eta=0 \). As expected, when \( N_E \) is small, the trajectories closely resemble those in pure gravitational sedimentation, with the limiting colliding trajectories reaching the collision sphere at \( \theta=\pi/2 \) (see \citealt{dhanasekaran2021collision}). The open and colliding trajectories in the background represent a typical trajectory map driven solely by differential sedimentation.  

In contrast, at large \( N_E \), the electric-field-induced attractive force enables surface-to-surface contact even when the initial horizontal separation between the droplet centers is significantly greater than in the weak-field case. Additionally, the forces and torques exerted by the strong electric field cause the trajectory to take a sharp bend before reaching the collision surface. Since gravity dominates the relative motion at large separations, far-field trajectories remain nearly indistinguishable from the open trajectories (solid blue lines in Figure \ref{Pair_trajectories_weak_and_strong_field}) that would occur in the absence of an electric field. However, when the electric field fully governs the dynamics, the relative trajectories both originate and terminate on the collision sphere, forming what we refer to as loop trajectories. Detailed calculations of these loop trajectories can be found in \citet{thiruvenkadam2023pair}. A set of these trajectories, shown as light black lines in Figure \ref{Pair_trajectories_weak_and_strong_field}, represents cases where a purely vertical electric field dictates the motion. In the near field, electric-field-induced forces and torques dominate, causing the limiting colliding trajectory to merge with a loop trajectory before reaching the collision surface. We identify a characteristic radial location, denoted as \( r_c \), where the effects of gravity and the electric field become comparable, resulting in a sharp trajectory bend. Later in this section, we use the scaling of \( r_c \) with \( N_E \) to predict the scaling behavior of collision efficiency in the strong-field regime.

\begin{figure*}
\centering
\includegraphics[width=1.0\textwidth]{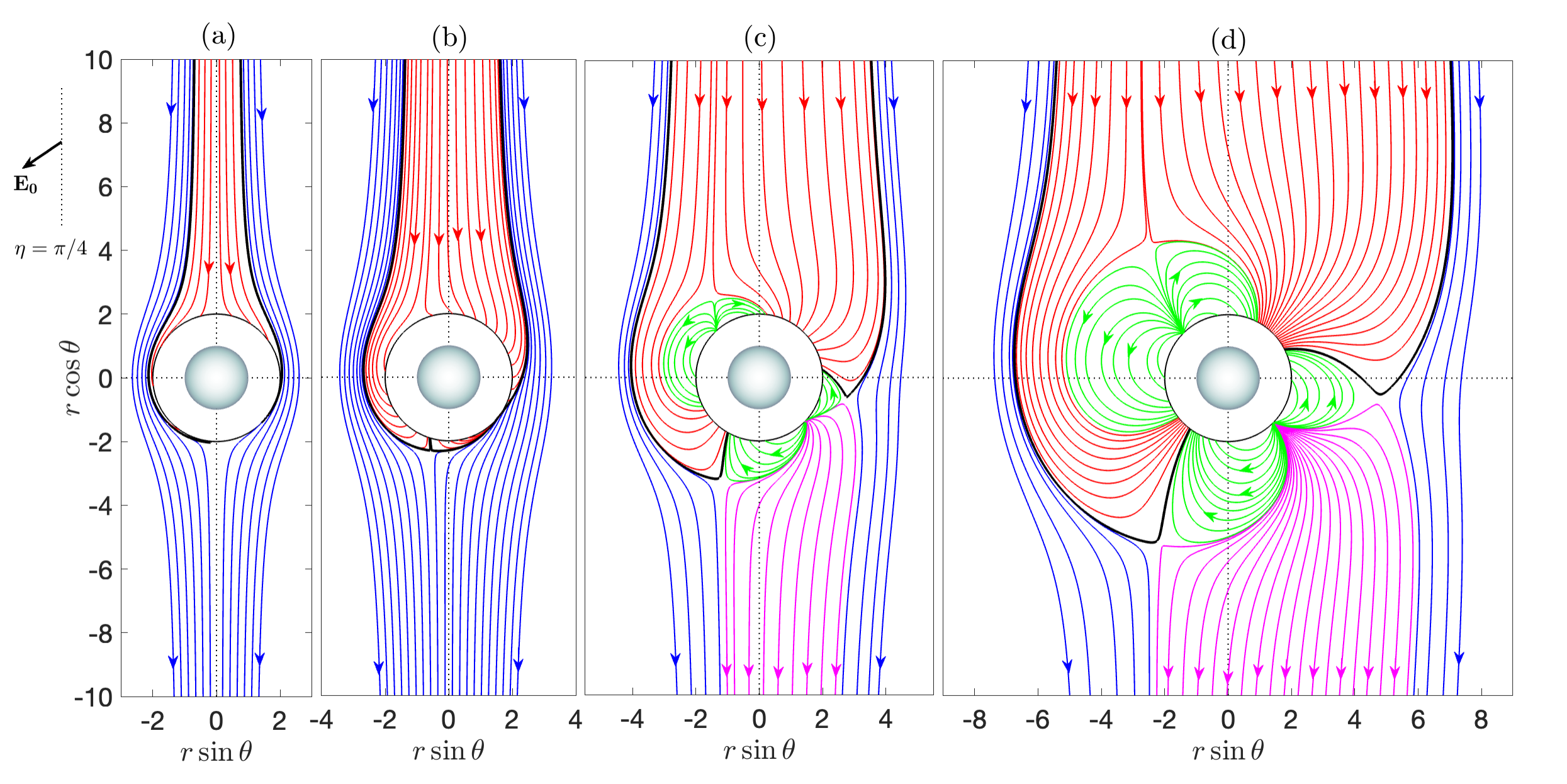}
\caption{Same as figure \ref{Pair_trajectories_electric_field_alpha_0} except $\eta=\pi/4$}
\label{Pair_trajectories_electric_field_alpha_45}
\end{figure*}

\begin{figure*}
\centering
\includegraphics[width=1.0\textwidth]{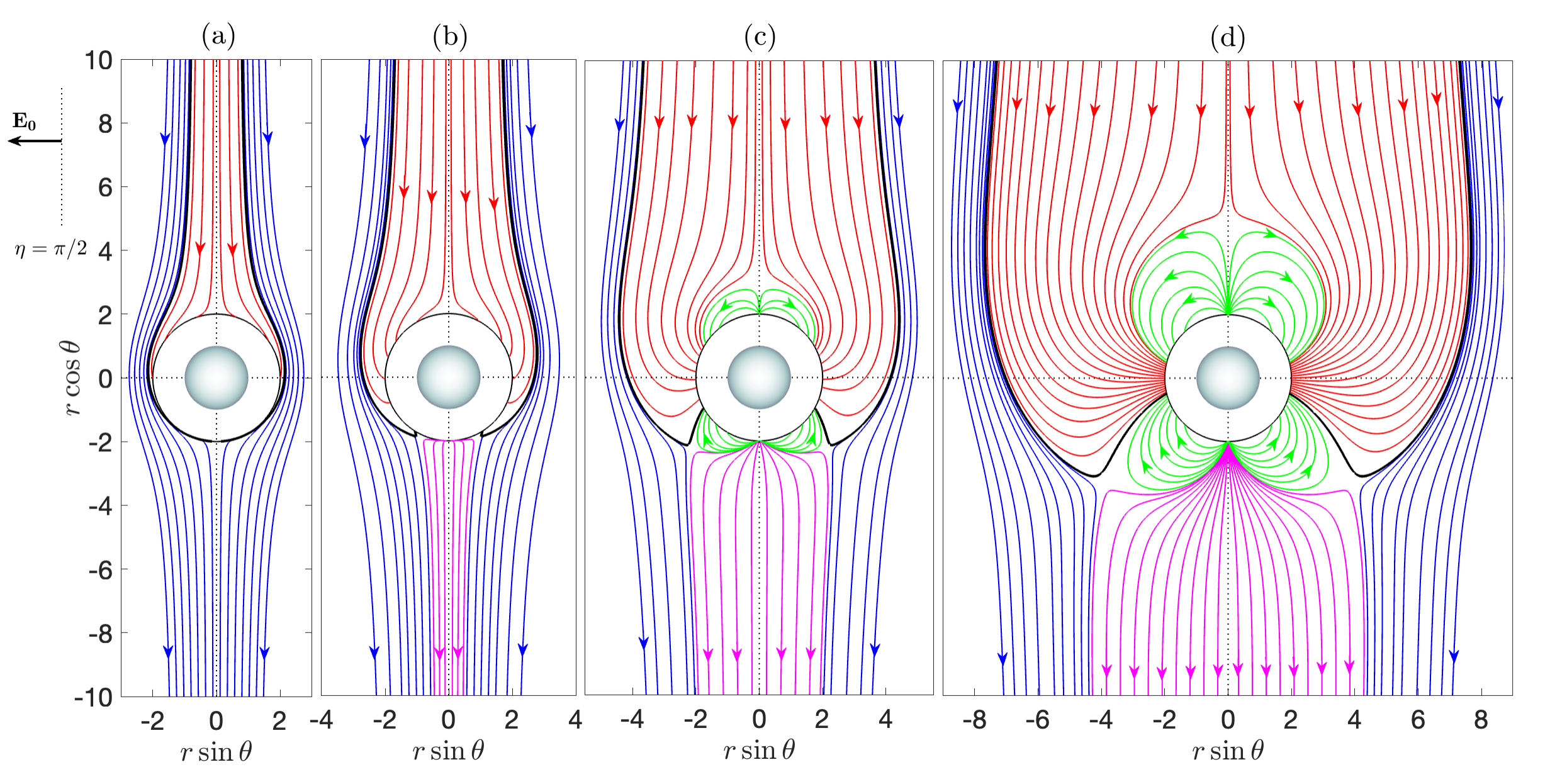}
\caption{Same as figure \ref{Pair_trajectories_electric_field_alpha_0} except $\eta=\pi/2$}
\label{Pair_trajectories_electric_field_alpha_90}
\end{figure*}

A detailed examination of how relative trajectory topologies evolve with increasing electric field strength is essential for accurately characterizing the collision dynamics. Figures \ref{Pair_trajectories_electric_field_alpha_0}, \ref{Pair_trajectories_electric_field_alpha_45}, and \ref{Pair_trajectories_electric_field_alpha_90} illustrate representative pair trajectories for three different electric field orientations: \( \eta = 0 \), \( \eta = \pi/4 \), and \( \eta = \pi/2 \), respectively, while keeping all other parameters constant. Specifically, we set \( \kappa = 0.5 \), \( Kn = 10^{-2} \), and \( N_v = 0 \), and examine the relative trajectories for four different electric field strengths: (a) \( N_E = 10^{-1} \), (b) \( N_E = 1 \), (c) \( N_E = 10 \), and (d) \( N_E = 10^2 \).  

For small \( N_E \), the relative trajectories resemble those observed in the absence of an electric field, consisting solely of open and colliding trajectories. However, as \( N_E \) increases, loop trajectories—characteristic of motion governed primarily by the electric field—begin to emerge. The locations on the collision surface where these loop trajectories originate depend on the angle \( \eta \) between gravity and the electric field. We find that loop trajectories initiate at \( \theta = \eta + (\pi/2) \) and \( \theta = \eta + (3\pi/2) \) and terminate at different locations on the collision sphere. These starting locations act as unstable fixed points in the system, but not all trajectories originating from these points form loops. Interestingly, some trajectories (shown in pink) diverge to infinity instead, and the volume occupied by these trajectories increases with increasing \( N_E \).  

As \( N_E \) grows asymptotically large, all colliding trajectories converge toward impact locations centered around \( \theta = \eta \) and \( \theta = \eta + \pi \). The limiting colliding trajectories serve as separatrices, depicted as thick black lines in Figures \ref{Pair_trajectories_electric_field_alpha_0}, \ref{Pair_trajectories_electric_field_alpha_45}, and \ref{Pair_trajectories_electric_field_alpha_90}. These separatrices distinguish open from colliding trajectories for small to moderate \( N_E \). However, at larger \( N_E \), they instead delineate colliding trajectories from open trajectories in the far field and from loop trajectories in the near field. Though not explicitly shown in the figures, for sufficiently large \( N_E \), the limiting colliding trajectories and outermost pink trajectories intersect at points of maximum curvature for both curves. These intersections correspond to saddle points in the system, whose positions are strongly influenced by \( \eta \).  

Despite this dependence on \( \eta \), the critical impact parameters consistently increase with increasing electric field strength. Additionally, the trajectories exhibit asymmetry about the gravity axis, except in cases where the electric field is oriented either purely vertically or horizontally. Importantly, the fundamental trajectory topology for strong electric fields remains unchanged even when droplets interact through pure continuum hydrodynamics. This is because the electric-field-induced forces can overcome continuum lubrication resistance, allowing droplets to make surface-to-surface contact within a finite time.

\begin{figure*}
\centering
\includegraphics[width=1.0\textwidth]{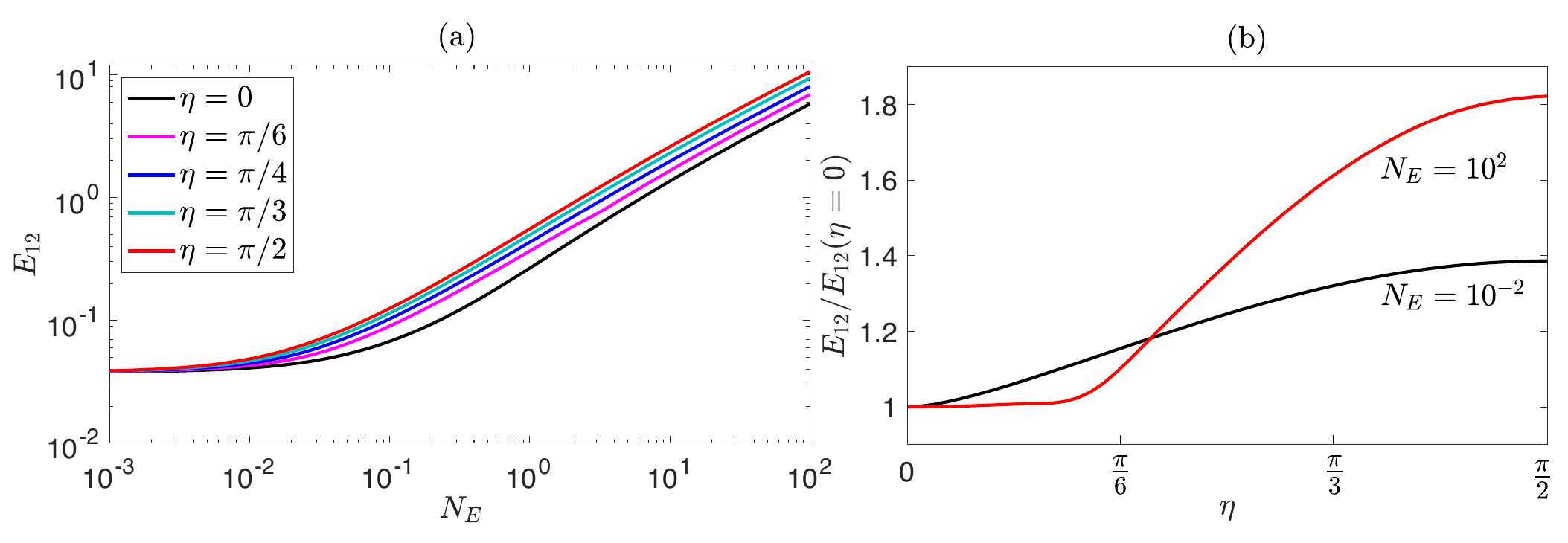}
\caption{(a) Collision efficiency as a function of the relative strength of electric-field-induced forces and gravity ($N_E$) for different angles between the electric field and gravity $\eta$ when $\kappa = 0.5$, $Kn = 10^{-2}$, and $N_v = 0$. (b) Collision efficiencies scaled with their values for $\eta=0$ as a function of $\eta$ for $N_E = 10^{-2}, 10^2$ when $\kappa = 0.8$, $Kn = 10^{-2}$, and $N_v = 0$.}
\label{Collision_efficiency_noncontinuum_without_vdW_with_different_eta}
\end{figure*}

\begin{figure*}
\centering
\includegraphics[width=1.0\textwidth]{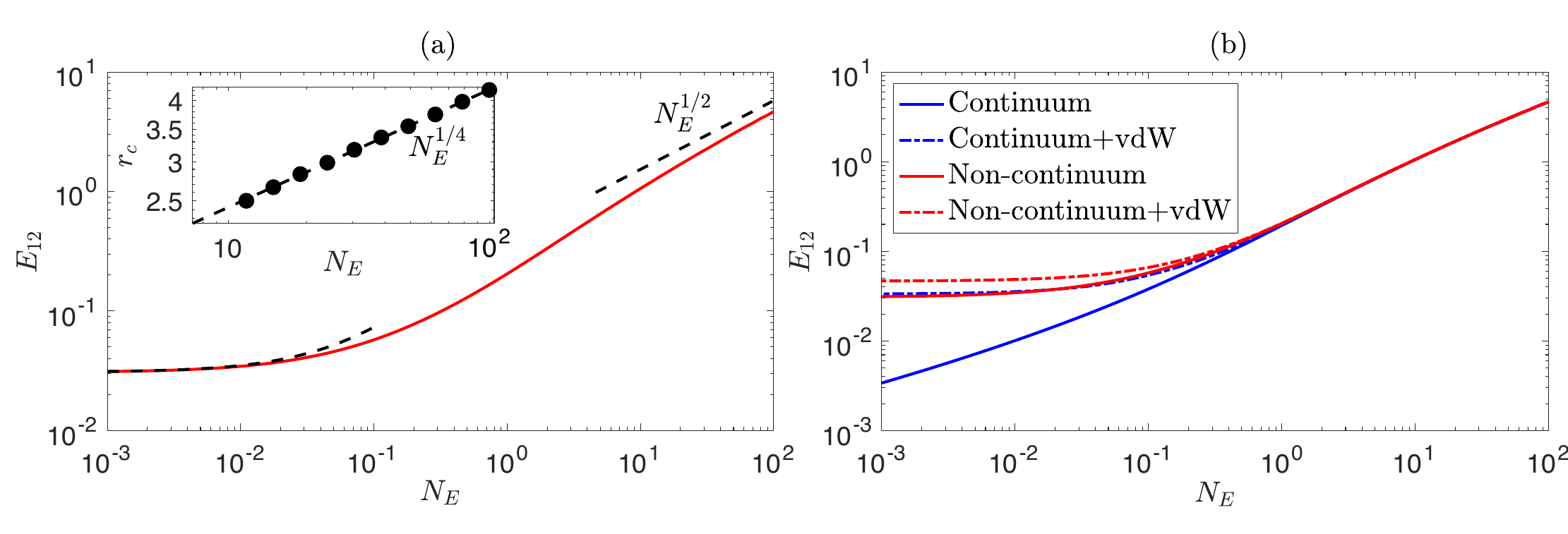}
\caption{(a) Collision efficiency as a function of $N_E$ with asymptotics for small and large $N_E$ limits when $\kappa=0.9$, $Kn=10^{-2}$, $N_v=0$, and $\eta=0$. The black dashed lines indicate the asymptotics. The inset shows the behavior of $r_c$ for $N_E \gg 1$. (b) Variation of collision efficiency with $N_E$ for different collision-inducing mechanisms. All parameters are the same as in (a) except for the cases with der Waals force where $N_L = 500$, and $N_v = 10^{-3}$.}
\label{Collision_efficiency_with_NE_asymptotics_and_different_mechanisms}
\end{figure*}

\begin{figure*}
\centering
\includegraphics[width=1.0\textwidth]{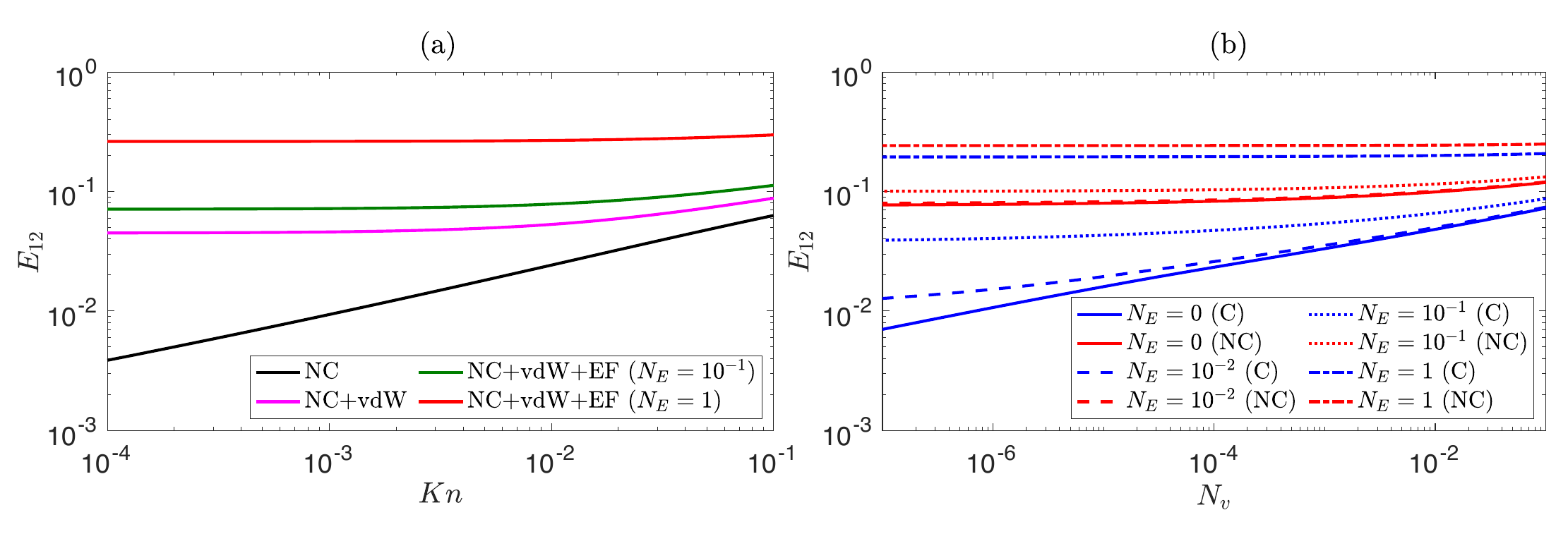}
\caption{Collision efficiency as a function of (a) $Kn$ for non-continuum (NC) hydrodynamics alone, non-continuum plus van der Waals (NC+vdW) interactions, and non-continuum hydrodynamics plus van der Waals plus electric-field-induced forces (NC+vdW+EF) when $\kappa = 0.5$, $N_L = 250$, $N_v = 10^{-2}$, and $\eta=0$ and (b) $N_v$ for both continuum (C) and non-continuum ($Kn=10^{-1}$) lubrication interactions when $\kappa = 0.9$, $N_L = 500$, $\eta=0$, and $N_E = 0, 10^{-2}, 10^{-1}, 1.0$.}
\label{Collision_efficiency_with_Kn_and_Nv}
\end{figure*}

We now investigate the effect of the electric field on collision efficiency, which we can determine from the trajectory analysis presented previously. The rest of this section will explore how collision efficiency varies with the different parameters involved in the problem. Figure \ref{Collision_efficiency_noncontinuum_without_vdW_with_different_eta}(a) illustrates the variation of collision efficiency with the relative strength of electric-field-induced force and gravity ($N_E$) for $\kappa = 0.5$, $Kn = 10^{-2}$, and $N_v = 0$ when the angles between the electric field and gravity are $\eta=0, \pi/6, \pi/4, \pi/3,$ and $\pi/2$. In the gravity-dominated regime (i.e., $N_E \ll 1$), all the curves tend to merge since the non-continuum lubrication interactions drive the collision mechanisms across all cases in a similar way for a given $\kappa$ and $Kn$. The collision efficiency increases monotonically with the increase in the electric field-induced force, and $E_{12}$ grows according to a power-law for large $N_E$. We find that the collision efficiency is higher for a higher $\eta$. To explicitly show the dependency of collision efficiency on $\eta$, we plot collision efficiency scaled by its value for $\eta=0$ as a function of $\eta$ for weak ($N_E=10^{-2}$) and strong ($N_E=10^2$) electric fields when $\kappa = 0.8$, $Kn = 10^{-2}$, and $N_v = 0$ (see figure \ref{Collision_efficiency_noncontinuum_without_vdW_with_different_eta}(b)). In both cases, the scaled collision efficiency increases as $\eta$ increases from $0$ to $\pi/2$. However, there is a subtle difference in how they increase with $\eta$. For a high $N_E$ value (i.e., $N_E=10^2$), the collision efficiency remains almost constant up to a certain value of $\eta$ and then increases monotonically with increasing $\eta$. This behavior has a correlation with the corresponding pair trajectory map in terms of the locations of the saddle points. We observe that both the saddle points lie above or below the $r\sin\theta =0$ line for $\eta$ values below or above this certain value. On the other hand, for a small value of $N_E$ (i.e., $N_E=10^{-2}$), the collision efficiency increases monotonically as $\eta$ increases, and this behavior corresponds to a small perturbation effect due to a weak electric field on the collision efficiency of two sedimenting droplets with non-continuum lubrication interactions. In the appendix, we derive the analytical expression for the collision efficiency up to $O(N_E)$ using a regular perturbation expansion. Figure \ref{Collision_efficiency_with_NE_asymptotics_and_different_mechanisms}(a) illustrates the validity of the asymptotic calculation in the small $N_E$ limit for a vertical electric field when $\kappa=0.9$, $Kn=10^{-2}$, and $N_v=0$. On the other hand, to predict the power-law behavior for large $N_E$, we present the following scaling argument. The first and second terms in the radial relative velocity equation (\ref{vr_equation}) are comparable at $r_c$, which typically happens to be a large separation. Now, $L, G \rightarrow 1$ and $F_1, F_2 \rightarrow 1/r^4$ in the far field. Thus, from equation (\ref{vr_equation}), we have $r_c \sim N_E^{1/4}$ and $E_{12} \sim r_c^2 \sim N_E^{1/2}$. This scaling relation is valid irrespective of the value of $\eta$. Figure \ref{Collision_efficiency_noncontinuum_without_vdW_with_different_eta}(a), where lines are parallel for large $N_E$, confirms this argument. We fit the data for large $N_E$ shown in the inset of figure \ref{Collision_efficiency_with_NE_asymptotics_and_different_mechanisms}(a), which confirms our scaling argument. Interestingly, electrostatic forces in strong electric fields dominate over other collision-inducing mechanisms, resulting in the scaling of collision efficiency remaining unaffected, even when considering full continuum hydrodynamics and the effects of van der Waals forces. (see figure \ref{Collision_efficiency_with_NE_asymptotics_and_different_mechanisms}(b)). The continuous blue line in figure \ref{Collision_efficiency_with_NE_asymptotics_and_different_mechanisms}(b) represents the collision efficiency when the droplet pair interacts via full continuum hydrodynamics. In this case, $E_{12}$ decreases rapidly with decreasing $N_E$, and theoretically, it will approach zero for $N_E \rightarrow 0$. However, in the presence of van der Waals force, the collision efficiency asymptotically approaches a finite value as $N_E \rightarrow 0$, even when droplets interact via continuum hydrodynamics (the dash-dotted blue line in figure \ref{Collision_efficiency_with_NE_asymptotics_and_different_mechanisms}(b)). This behavior arises because van der Waals forces drive the collision dynamics in the gravity-dominated regime when droplets interact through full continuum hydrodynamics. As expected, for small to moderate values of $N_E$, the collision efficiency due to non-continuum hydrodynamics plus van der Waals interactions (the dash-dotted red line in figure \ref{Collision_efficiency_with_NE_asymptotics_and_different_mechanisms}(b)) becomes higher than that due to non-continuum hydrodynamics alone. 

We will now demonstrate how the variation of the strength of non-continuum lubrication effects modifies the collision dynamics. Figure \ref{Collision_efficiency_with_Kn_and_Nv}(a) shows the collision efficiency as a function of the Knudsen number for $N_E = 10^{-1}, 1$ values when $\kappa=0.5$, $N_L=250$, $N_F=10^{-2}$, and $\eta=0$. We compare our findings with those of \citet{dhanasekaran2021collision}, who calculated the collision efficiency of two sedimenting droplets considering non-continuum hydrodynamics, both with and without the inclusion of van der Waals forces. The black line in figure \ref{Collision_efficiency_with_Kn_and_Nv}(a) represents the collision efficiency resulting from non-continuum effects (NC) alone, as calculated by \citet{dhanasekaran2021collision} through the evaluation of the integral in equation (\ref{collision_efficiency_due_to_hydrodynamics_only}) for various values of $Kn$. As $Kn$ decreases, the relative thickness of the non-continuum lubrication layer decreases, causing $E_{12}$ due to NC alone to decrease monotonically and approach zero in the $Kn \rightarrow 0$ limit. The collision efficiency due to the combined effects of non-continuum hydrodynamics and van der Waals forces (NC+vdW) also decreases with decreasing $Kn$ (see the pink line in figure \ref{Collision_efficiency_with_Kn_and_Nv}(a)). However, in this case, asymptotes to a finite value as $Kn$ approaches zero because van der Waals attraction forces dictate the asymptotic behavior for small values of $Kn$, while non-continuum effects become insignificant. Expectedly, including electric-field-induced forces alongside non-continuum and van der Waals interactions (i.e., NC+vdW+EF) increases the collision efficiencies compared to the previous two cases. The behavior of $E_{12}$ with $Kn$ is quite similar to the NC+vdW case when the electric field strength is relatively weak (e.g., $N_E = 10^{-1}$). However, when the electric field is strong, electric-field-induced forces dominate over the combined effects of NC and vdW, causing the collision efficiency to become independent of $Kn$. As a result, the $Kn$ vs. $E_{12}$ curve becomes a straight line parallel to the $Kn$ axis (see the red line for $N_E = 1$ in Figure \ref{Collision_efficiency_with_Kn_and_Nv}(a)). To investigate the behavior of collision efficiency under varying strengths of the van der Waals force, we plot $E_{12}$ for $N_v$ ranging from $10^{-7}$ (indicating a weak van der Waals force) to $10^{-1}$ (indicating a strong van der Waals force) with the parameters $\kappa = 0.9$, $Kn = 10^{-1}$, $N_L = 500$, and $N_E = 0, 10^{-2}, 10^{-1}, 1$ (see Figure \ref{Collision_efficiency_with_Kn_and_Nv}(b)). In this figure, we also compare the results to their continuum counterparts. As anticipated, the collision efficiency ($E_{12}$) decreases as the strength of the van der Waals force diminishes (i.e., as $N_v$ decreases). This decrease in $E_{12}$ becomes particularly pronounced when the field strength is zero ($N_E=0$) or low ($N_E=10^{-2}$), and droplets interact via continuum hydrodynamics. When we consider non-continuum lubrication interactions, $E_{12}$ decreases relatively slowly and asymptotically approaches the value corresponding to collision efficiency due to non-continuum effects alone at $N_E=0$ and to the value corresponding to non-continuum effects combined with electric-field-induced forces at $N_E=10^{-2}, 10^{-1}$. Interestingly, the collision efficiency remains independent of the van der Waals force when the imposed field strength is high (for instance, $N_E = 1$), as electric-field-induced forces dominate all other factors contributing to a finite collision rate.

\begin{figure}
\centering
\includegraphics[width=0.45\textwidth]{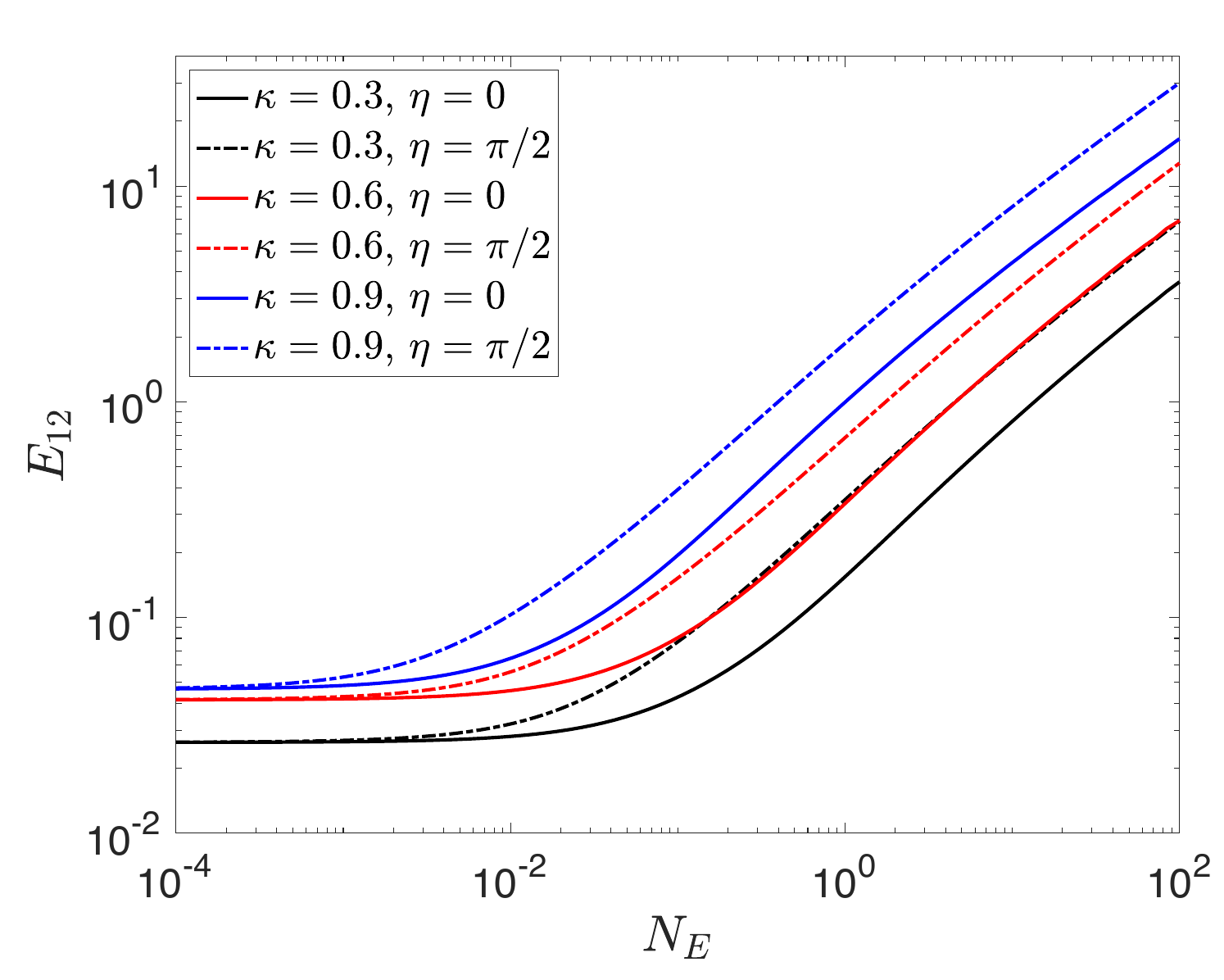}
\caption{Collision efficiency as a function of $N_E$ for Kn$=10^{-2}$, $N_L = 500$, and $N_v = 10^{-3}$ when $\kappa=0.3, 0.6,$ and $0.9$, and $\eta=0,\pi/2$.}
\label{Collision_efficiency_with_NE_kappa_03_05_09_NL_500_NF_1000_Kn_001}
\end{figure}

To examine how size differences between cloud droplets influence their dynamics under the combined effects of non-continuum hydrodynamics, van der Waals forces, and electric-field-induced forces, we plot the collision efficiency as a function of \(N_E\) for \(\kappa = 0.3, 0.6, 0.9\) and \(\eta = 0, \pi/2\) with \(Kn = 10^{-2}\), \(N_L = 500\), and \(N_v = 10^{-3}\) (see Figure \ref{Collision_efficiency_with_NE_kappa_03_05_09_NL_500_NF_1000_Kn_001}). The collision efficiency trends for each size ratio are consistent with those discussed earlier (Figures \ref{Collision_efficiency_noncontinuum_without_vdW_with_different_eta}(a) and \ref{Collision_efficiency_with_NE_asymptotics_and_different_mechanisms}(b)). In the gravity-dominated regime, collision efficiency decreases as \(\kappa\) decreases because the smaller droplet tends to follow the flow streamlines and move around the larger droplet. As a result, collisions occur only if the smaller droplet follows a streamline very close to the larger one. Additionally, as the size ratio decreases, electric-field-induced forces weaken, further reducing the collision efficiency across all values of \(N_E\).  

\begin{figure*}
\centering
\includegraphics[width=1.0\textwidth]{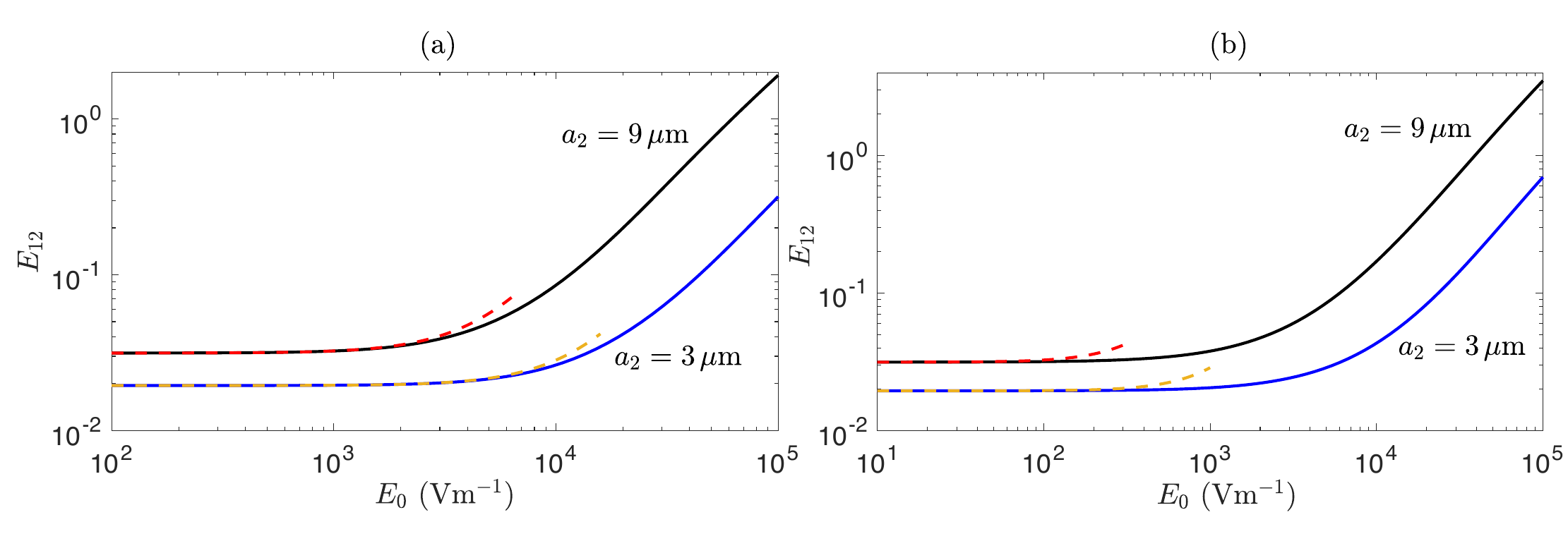}
\caption{Collision efficiency as a function of the strength of the external electric field for a pair of water droplets in air with $a_1=10$ \textmu m and $a_2 = 3, 9$ \textmu m when van der Waals forces are absent, and the electric field acts in the (a) vertical direction and (b) horizontal direction. Dashed lines are from the asymptotic expression of collision efficiency for the weak electric field.}
\label{Collision_efficiency_with_E0_without_vdW}
\end{figure*}

\begin{figure}
\centering
\includegraphics[width=0.45\textwidth]{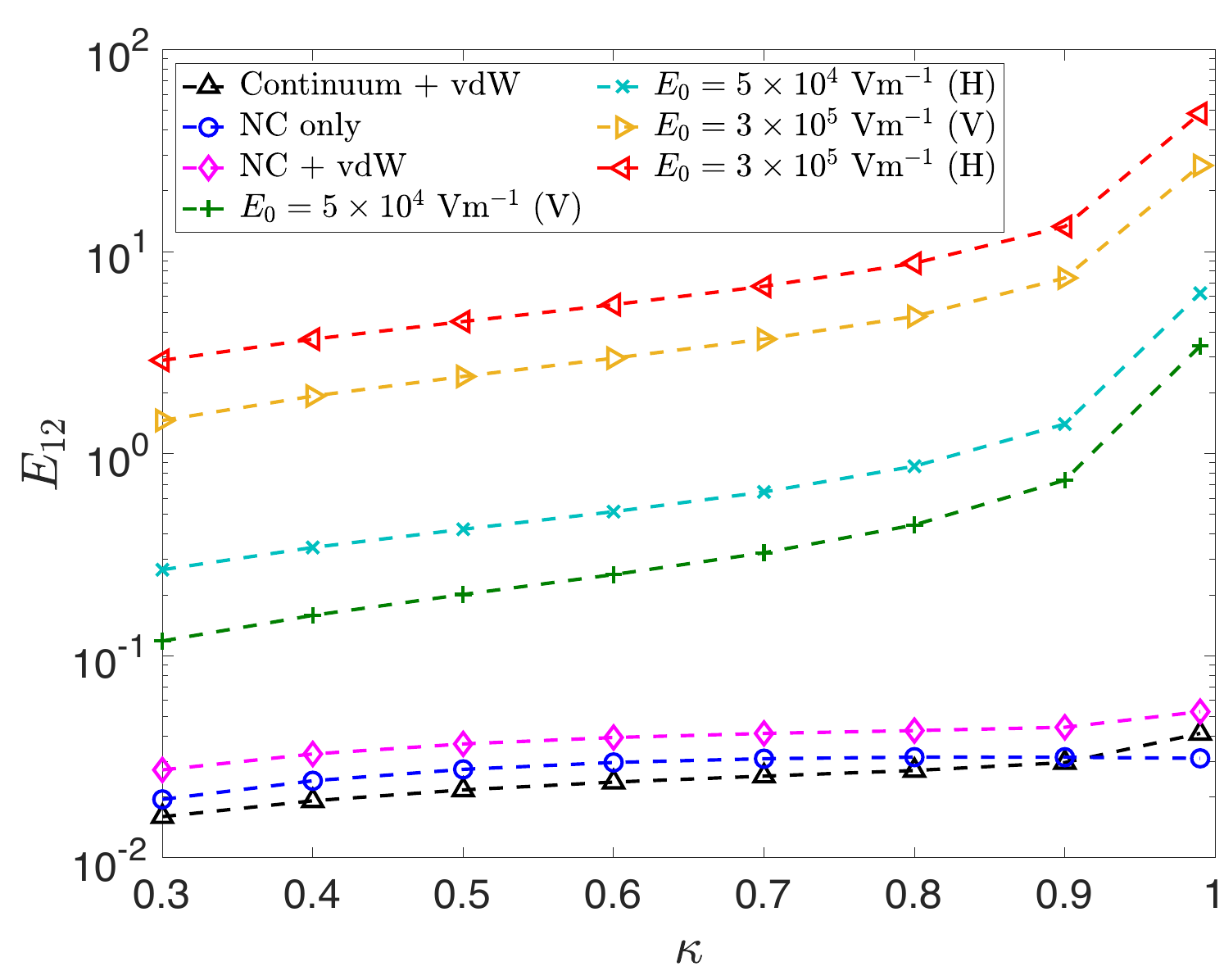}
\caption{Collision efficiency as a function of the size ratio for water droplets in air with $a_1 = 10$ \textmu m, vertical (indicated by ``V" within parentheses) and horizontal (indicated by ``H" within parentheses) electric field $E_0 = 5 \times 10^4$, $3 \times 10^5$ Vm$^{-1}$ when non-continuum effects (NC), van der Waals interactions (vdW) and electric-field-induced forces (EF) act together. We included results from previous studies that predicted collision efficiencies without an external electric field to compare our findings.}
\label{Collision_efficiency_a1_10}
\end{figure}

Our discussion thus far has focused on how collision efficiency varies with different nondimensional quantities relevant to the problem. Now, we will present some results considering the property values of water droplets under typical cloud conditions when the radius of the larger droplet $a_1 = 10$ \textmu m. Typical values for droplet density ($\rho_p$), air density ($\rho_f$), dynamic viscosity of air ($\mu_f$), and electric field strength ($E_0$) are provided in Sec. \ref{Introduction}. The mean free path of air increases with altitude, and for warm cumulus clouds, $\lambda_0 \sim 0.1$ \textmu m (see \citealt{wallace2006atmospheric}). Consequently, we express the Knudsen number as a function of the size ratio as $Kn = 0.02/(1+\kappa)$. For water droplets in air, the Hamaker constant is $A_H \approx 3.7 \times 10^{-20}$ J (see \citealt{friendlander2000smoke}). Therefore, we express the parameters $N_L$ and $N_v$ in terms of $\kappa$ as follows: $N_L = 6.28 \times 10^2 (1+\kappa)$ and $N_v = 1.77 \times 10^{-4}/[\kappa (1-\kappa^2)]$. The parameter $N_E$ defined above does not depend on the size ratio, and it varies with the strength of the electric field according to the relation $N_E= 2.66 \times 10^{-10} E_0^2$. For a given $\kappa$ and $E_0$, we determine the required dimensionless parameters ($Kn$, $N_L$, $N_v$, and $N_E$) from these relations and then calculate the collision efficiency. Figure \ref{Collision_efficiency_with_E0_without_vdW} shows how collision efficiency increases as the strength of the vertical or horizontal electric field increases when $a_2 = 3$ \textmu m (i.e., $\kappa = 0.3$ indicating droplets with significantly different in size) and $a_2 = 9$ \textmu m (i.e., $\kappa = 0.9$ indicating droplets with almost equal in size). For electric field strengths $E_0$ up to a few thousand Vm$^{-1}$, $E_{12}$ increases very slowly, suggesting that the contribution of the fair-weather electric field in droplet-droplet collisions is insignificant. The dashed lines in figure \ref{Collision_efficiency_with_E0_without_vdW} represent results calculated from the analytical expression (\ref{Analytical_expression_weak_field}) for the weak electric fields. These results also demonstrate that the asymptotic expression accurately predicts the collision efficiency influenced by the fair-weather atmospheric electric field. The enhancement in collision efficiency becomes rapid when the electric field strengths exceed $10^4$ Vm$^{-1}$, typical in strongly electrified clouds. More precisely, for strong electric fields, $E_{12} \sim N_E^{1/2} \sim E_0$ since $N_E$ varies with the square of the strength of the electric field.

Finally, we estimate how collision efficiency varies with the size ratio $\kappa$ of cloud droplets for a vertical or horizontal electric field of strengths $5 \times 10^4$ Vm$^{-1}$ and $3 \times 10^5$ Vm$^{-1}$ and in the presence of van der Waals forces (see figure \ref{Collision_efficiency_a1_10}). Here also, we keep $a_1 = 10$ \textmu m, and thus, the above relationships of the relevant parameters with $\kappa$ and $E_0$ are applicable. The collision efficiency increases with the size ratio as the electric-field-induced forces strengthen with a higher size ratio. We report that for a given droplet pair, the collision efficiency increases by an order of magnitude when vertical or horizontal electric field strength increases from $5 \times 10^4$ Vm$^{-1}$ to $3 \times 10^5$ Vm$^{-1}$. These findings also indicate that a horizontal electric field is more effective than a vertical one in promoting collisions between droplets. As discussed in \S \ref{Introduction}, gravity-induced collisions due to continuum and non-continuum hydrodynamics and van der Waals forces have extensive treatment in the literature (see \citet{davis1984rate,dhanasekaran2021collision}). The comparison of these results (three bottom lines in figure \ref{Collision_efficiency_a1_10} represent these results) with our present findings demonstrates that an external electric field always enhances collision efficiency.

\section{Summary and conclusions}\label{Conclusions}

We have quantified the influence of an external electric field on the collision rate of uncharged cloud droplets settling under gravity in a quiescent atmosphere. Our study builds upon recent analytical formulations for electric-field-induced forces at close separations, demonstrating that these forces can overcome lubrication resistance and enable surface-to-surface contact within a finite time. Additionally, we incorporated non-continuum hydrodynamic effects, which are essential for accurately modeling cloud droplet collisions. The pair trajectory maps presented in Figures \ref{Pair_trajectories_electric_field_alpha_0}, \ref{Pair_trajectories_electric_field_alpha_45}, and \ref{Pair_trajectories_electric_field_alpha_90} illustrate how electric field-induced forces modify droplet trajectories compared to purely gravity-driven interactions. Our findings indicate that electric-field-induced forces and non-continuum hydrodynamic effects collectively enhance droplet collisions in clouds.   

While our analysis provides new insights, it does not account for the effects of droplet inertia, which can significantly influence the collision rate for larger droplet pairs. Moreover, cloud droplets often carry surface charges, making it necessary to consider both external electric field effects and direct electrostatic interactions between charged droplets. Our recent studies \citep{patra2023collision} on like-charged settling droplets (neglecting inertia) have shown a non-monotonic relationship between surface charge and collision efficiency: an initial increase due to near-field attraction, followed by a sharp decrease due to far-field repulsion. A natural extension of our work would be to incorporate exact hydrodynamic and electrostatic forces into the collision rate calculations for sedimenting inertial droplets, which may reveal further complexities in droplet interactions.  

Another crucial factor not addressed in this study is atmospheric turbulence, which plays a key role in the growth of cloud droplets and the initiation of rainfall \citep{shaw2003particle}. In the size-gap regime, turbulence enhances collision rates through (i) increased radial relative velocities between droplet pairs \citep{saffman1956collision,falkovich2007sling} and (ii) preferential concentration of inertial droplets in straining regions of the flow \citep{sundaram1997collision,chun2005clustering}. Additionally, the velocity perturbations induced by droplet motion further influence collision rates under realistic conditions \citep{pinsky2007collisions}. Recent direct numerical simulations coupled with Lagrangian particle tracking \citep{chen2018turbulence, michel2023influence} have demonstrated that turbulence and droplet aerodynamic interactions significantly broaden the droplet size distribution. However, there are currently no theoretical predictions for collision efficiencies of inertial droplets under the combined effects of turbulence, gravity, and an external electric field. Future studies should aim to incorporate these factors into collision rate calculations, including a rigorous treatment of short-range electric-field-induced forces.  

Our discussion has focused exclusively on warm clouds, where droplet collisions govern the formation of raindrops. However, mixed-phase clouds contain both supercooled droplets and ice crystals, and their interactions play a fundamental role in precipitation formation. For instance, snow aggregates form through collisions between ice crystals, while graupel growth occurs via rimming - when settling ice crystals collide with supercooled droplets in turbulent conditions \citep{pruppacher1997microphysics,wang1994collision}. While spherical droplet collisions are well understood, ice crystal collisions remain less explored due to their anisotropic shapes and variable settling orientations. Recent studies \citep{jucha2018settling,sheikh2022colliding} have examined turbulence and gravitational effects on ice crystal collisions, while others have explored collisions between ice crystals and supercooled droplets \citep{naso2018collision,jost2019effect,sheikh2024effect}. However, these studies rely on the ghost collision approximation, which neglects hydrodynamic and electrostatic interactions between colliding hydrometeors. Given that electrostatic forces can become significant at close distances, they may substantially influence ice crystal collision outcomes. Recent work by \citet{joshi2025electrostatic} has quantified electrostatic forces and torques between charged anisotropic particles, showing that electrostatic torques can induce preferential alignment of ice crystals, thereby affecting their interactions. Future research should incorporate these electrostatic effects into collision rate calculations, considering both translational and rotational dynamics to achieve a more comprehensive understanding of hydrometeor interactions due to electric effects.  


\acknowledgments
The authors acknowledge support from IIT Madras for its support of the ‘Laboratory for Atmospheric and Climate Sciences’ research initiative under the Institute of Eminence framework. P.P. would like to acknowledge financial support from the Prime Minister’s Research Fellows (PMRF) scheme, Ministry of Education, Government of India (Project Number: SB22230184AMPMRF008746)


\datastatement
The data that support the findings of this study are available
within the article.


\appendix
\appendixtitle{Analytical expression for the collision efficiency when $N_E \ll 1$ and $N_v = 0$} \label{appendixA}

We can analytically determine the relative trajectories when the applied electric field is weak. Let's assume the solution for $\theta(r)$ takes the form $\theta(r) = \theta_0(r) + N_E \theta_1(r) + O(N_E^2)$, where $\theta_0$ and $\theta_1$ represent the solutions at $O(1)$ and $O(N_E)$, respectively. In the absence of the van der Waals force ($N_v = 0$), the relative trajectory equations at $O(1)$ and $O(N_E)$, derived from equation (\ref{final_trajectory_equation}), are as follows:
\begin{gather}
\frac{d \theta_0}{d r} = -\frac{M}{r L}\tan\theta_0, \hspace{3mm} \text{and} \label{O(1)_trajectory_equation} \\
\frac{d \theta_1}{d r} +  \frac{M \sec^2\theta_0}{r L}\theta_1 = Q(r), \label{O(NE)_trajectory_equation}
\end{gather}
where
\begin{gather}
Q(r) = \frac{G M \sin\theta_0}{r L^2 \cos^2\theta_0} C_{\kappa}  
\left(F_1\cos^2(\theta_0-\eta)+F_2\sin^2(\theta_0 -\eta)\right) \nonumber \\ + \frac{H \sec\theta_0}{r L} C_{\kappa} F_8 \sin 2(\eta-\theta_0). 
\end{gather}
Here, $C_{\kappa} = \kappa/(1-\kappa)$. We need to find the dimensionless critical impact parameter to calculate the collision efficiency by determining the limiting colliding trajectory. To find the limiting colliding trajectory we set $v_r = 0$ and then obtain the boundary conditions for (\ref{O(1)_trajectory_equation}) and (\ref{O(NE)_trajectory_equation}). Thus, the boundary conditions for the $O(1)$ and $O(N_E)$ trajectory equations are as follows:
\begin{gather}
\theta_0 (r = 2) = \frac{\pi}{2}, \hspace{3mm} \text{and} \label{O(1)_boundary_condition} \\
\theta_1 = \left.\frac{G C_{\kappa} \left(F_1\cos^2(\theta_0(r)-\eta)+F_2\sin^2(\theta_0(r)-\eta)\right)}{L \sin\theta_0(r)}\right\vert_{r=2} \nonumber \\ = B \hspace{3mm} \text{(let's say).} \label{O(NE)_boundary_condition}
\end{gather}
The solutions for $\theta_0(r)$ and $\theta_1(r)$ are
\begin{gather}
\theta_0(r) = \sin^{-1}\Bigg[\exp\left(\int_2^r-\frac{M(r')}{r' L(r')}dr'\right)\Bigg], \label{Expression_for_theta0}\\
\theta_1(r) = \exp \left(-\int_2^r \frac{M(r') \sec^2\theta_0(r')}{r' L(r')} dr'\right) \Bigg[B + \nonumber \\ \int_2^r Q(r') \exp \left(\int_2^{r'} \frac{M(r'') \sec^2\theta_0(r'')}{r'' L(r'')} dr''\right)dr' \Bigg]. \label{Expression_for_theta1} 
\end{gather}
For a weak electric field, the upstream interception area is approximately a circle, and thus, $\overline{y}_{c+} \approx  \overline{y}_{c-}=\overline{y}_{c}$. The dimensionless critical impact parameter can be written as 
\begin{gather}
\overline{y}_{c} = \lim_{r \rightarrow \infty}r\sin\theta = \lim_{r \rightarrow \infty}r\sin \left(\theta_0(r) + N_E \theta_1(r)\right) + O(N_E^2) \nonumber \\ = \overline{y}_{c0} + N_E \overline{y}_{c1} + O(N_E^2). \label{yc_expression}
\end{gather}
In equation (\ref{yc_expression}), $\overline{y}_{c0}$ and $\overline{y}_{c1}$ can be expressed as
\begin{gather}
\overline{y}_{c0} = \lim_{r \rightarrow \infty}r\sin\theta_0 = 2\exp\left(-\int_2^{\infty}\frac{M-L}{r L}dr\right), \label{yc0_expression} \\
\hspace{-38mm} \overline{y}_{c1} = \lim_{r \rightarrow \infty}(r\cos\theta_0)\theta_1 \nonumber \\ = \lim_{r \rightarrow \infty}(r\sin\theta_0)  \lim_{r \rightarrow \infty}\left(\dfrac{1}{\sin^2\theta_0(r)}-1\right)^{1/2} \theta_1(r) \nonumber \\ \nonumber \\ = \overline{y}_{c0} \left(\exp\left(2\int_2^{\infty}\frac{M}{r L}dr\right)-1\right)^{1/2} \lim_{r \rightarrow \infty} \theta_1(r). \label{yc1_expression}
\end{gather}
The expression for the collision efficiency in terms of $\overline{y}_{c0}$ and $\overline{y}_{c1}$ becomes
\begin{gather}
E_{12} = \frac{1}{4}\overline{y}_c^2 = \frac{1}{4}\left(\overline{y}_{c0}^2+ 2 N_E \overline{y}_{c0}\overline{y}_{c1}\right) + O(N_E^2).
\label{Collision_efficiency_small_NE}
\end{gather}
Thus, the final expression for the collision efficiency up to $O(N_E)$ becomes
\begin{gather}
E_{12} = \exp\left(-2\int_2^{\infty}\frac{M-L}{r L}dr\right) \Bigg(1+ 2 N_E   \nonumber \\ \Bigg[\exp\left(2\int_2^{\infty}\frac{M}{r L}dr\right)-1\Bigg]^{1/2} \lim_{r\rightarrow\infty}\theta_1(r)\Bigg) + O(N_E^2)
\label{Analytical_expression_weak_field}
\end{gather}


\bibliographystyle{ametsocV6}
\bibliography{references}

\end{document}